\documentclass[conference]{IEEEtran}
\IEEEoverridecommandlockouts

\usepackage[utf8]{inputenc}
\usepackage{graphicx}
\usepackage{textcomp}
\usepackage{booktabs}
\usepackage{amsmath}
\usepackage{amsfonts}
\usepackage{amsthm}
\usepackage{algorithm}
\usepackage{algpseudocode}
\usepackage{csquotes}
\usepackage{enumitem}
\usepackage{multirow}
\usepackage{multicol}
\usepackage{tabularx}
\usepackage{array}
\usepackage[dvipsnames]{xcolor}
\definecolor{mygreen}{rgb}{0.0, 0.5, 0.0}
\usepackage{subcaption}
\usepackage{flushend}
\usepackage{comment}
\usepackage{mathrsfs}
\usepackage{gensymb}
\usepackage{bbold}
\usepackage{mathtools}
\newtheorem{definition}{Definition}

\newtheorem{example}{Example}
\newtheorem{remark}{Remark}
\usepackage{hyperref} 
\usepackage{tcolorbox}
\usepackage{wasysym}

\usepackage{framed}
\renewenvironment{leftbar}[1][\hsize]
{%
  \MakeFramed{\hsize#1\advance\hsize-\width\FrameRestore\vskip-3pt}
}
{\vskip-3pt\endMakeFramed} 
\def \calM{{\mathcal{M}}}
\def \calA{{\mathcal{A}}}
\def \calX{{\mathcal{X}}}
\def \calY{{\mathcal{Y}}}
\def \eg{\emph{e.g.}}
\def \ie{\emph{i.e.}}

\def \bff{\mathbf{f}}

\def \pstar{p^*}
\def \qstar{q^*}
\def \setOne{\mathbb{1}}
\def \eeps{e^{\varepsilon}}
\def \vs{\emph{vs.}}
\def\BibTeX{{\rm B\kern-.05em{\sc i\kern-.025em b}\kern-.08em
    T\kern-.1667em\lower.7ex\hbox{E}\kern-.125emX}}
\begin{document}

\title{\huge Revisiting Locally Differentially Private Protocols: Towards Better Trade-offs in Privacy, Utility, and Attack Resistance}

\author{\IEEEauthorblockN{Héber H. Arcolezi}
\IEEEauthorblockA{\textit{Inria}, Grenoble, France \\
\textit{ÉTS Montréal}, Montréal, Canada \\
heber.hwang-arcolezi@etsmtl.ca}
\and
\IEEEauthorblockN{Sébastien Gambs}
\IEEEauthorblockA{\textit{Université du Québec à Montréal (UQAM)} \\
Montréal, Canada \\
gambs.sebastien@uqam.ca}
}

\maketitle

\begin{abstract}
Local Differential Privacy (LDP) offers strong privacy protection, especially in settings in which the server collecting the data is untrusted. However, designing LDP mechanisms that achieve an optimal trade-off between privacy, utility and robustness to adversarial inference and integrity attacks remains challenging. In this work, we introduce a general multi-objective optimization framework for refining LDP protocols, enabling the joint optimization of privacy and utility under various adversarial settings. While our framework is flexible to accommodate multiple privacy and security attacks as well as utility metrics, in this paper, we specifically optimize for Attacker Success Rate (ASR) under \emph{data reconstruction attack} as a concrete measure of privacy leakage and Mean Squared Error (MSE) as a measure of utility. Complementarily, we evaluate integrity-oriented threats through data poisoning attacks, providing an additional adversarial perspective. More precisely, we systematically revisit these trade-offs by analyzing eight state-of-the-art LDP frequency estimation protocols and proposing refined counterparts that leverage tailored optimization techniques. Experimental results demonstrate that our proposed adaptive mechanisms consistently outperform their non-adaptive counterparts, achieving substantial reductions in ASR while preserving utility, and pushing closer to the ASR-MSE Pareto frontier. By bridging the gap between theoretical guarantees and real-world vulnerabilities, our framework enables modular and context-aware deployment of LDP mechanisms with tunable privacy-utility-attackability trade-offs.
\end{abstract}


\section{Introduction} \label{sec:introduction}

Differential Privacy (DP)~\cite{Dwork2006} has become a gold standard for preserving privacy in data analytics and mining, in which the goal is to ensure that an individual’s data does not significantly influence the output of any analysis. 
However, traditional DP relies on a trusted aggregator to apply the privacy mechanism, which is often impractical for decentralized or privacy-sensitive applications. 
This limitation has led to the rise of Local Differential Privacy (LDP)~\cite{first_ldp}, a model that removes the need for a trusted aggregator by requiring users to perturb their data locally before sharing it with the server.

The local DP model has gained widespread adoption, with major technology companies integrating it into their systems to enhance user privacy. 
Notable examples include Google Chrome~\cite{rappor}, Apple iOS~\cite{apple} and Windows 10~\cite{microsoft}, in which LDP protocols have been used to collect statistics while safeguarding individual data.
A fundamental task under LDP guarantees is \textbf{frequency estimation}, which forms the basis of many advanced data analysis tasks like heavy hitter estimation~\cite{Bassily2015,Li2024,Zhang2025}, frequency monitoring~\cite{rappor,microsoft,Arcolezi2022,Arcolezi2023evolving}, multidimensional queries~\cite{Gu2019,yang2020answering,Arcolezi_rs_fd,Filho2023}, frequent item-set mining~\cite{Li2022,Wu2023} and spatial density estimation~\cite{alptekin2023building,Tire2024}.

Due to its importance, numerous LDP frequency estimation protocols have been proposed~\cite{Raab2025}, namely, Generalized Randomized Response (GRR)~\cite{kairouz2016discrete}, Subset Selection (SS)~\cite{wang2016mutual,Min2018}, Symmetric Unary Encoding (SUE)~\cite{rappor}, Optimized Unary Encoding (OUE)~\cite{tianhao2017}, Summation with Histogram Encoding (SHE)~\cite{Dwork2006}, Thresholding with Histogram Encoding (THE)~\cite{tianhao2017}, Binary Local Hashing (BLH)~\cite{Bassily2015} and Optimal Local Hashing (OLH)~\cite{tianhao2017}. 
These protocols mainly focus on improving utility, often quantified in terms of the variance (\ie, Mean Squared Error -- MSE), as well as optimizing computational and communication costs~\cite{Hadamard,Feldman2022,arcolezi2025private} to enable efficient data collection in large-scale systems.

While utility and communication costs have been the traditional focus of LDP frequency estimation protocols, recent research has shed light on their vulnerabilities in adversarial settings. 
For instance, re-identification risks~\cite{Murakami2021, Arcolezi2023} have shown that adversaries can uniquely identify users within a dataset, while attribute inference attacks~\cite{Gadotti2022,Gursoy2022,Arcolezi2023evolving} have been explored in the context of iterative data collections.
Furthermore, recent works have examined data reconstruction attacks~\cite{Gursoy2022,Arcolezi2023,arcolezi2024revealing}, in which adversaries aim to recover the user's original input from its obfuscated output.
Among these threats, we focus on particularly critical reconstruction attacks as they represent a worst-case leakage scenario: if an attacker can recover the user's data from a single report, other adversaries with access to side information or repeated queries can only perform better (Examples~\ref{ex:re_identification} and \ref{ex:attribute_inference}).

\textbf{Contributions.} 
This paper introduces a \emph{novel attack-aware evaluation framework} for analyzing and improving LDP frequency estimation protocols, designed to capture the joint trade-off between privacy leakage and utility under adversarial settings. 
Our approach evaluates LDP mechanisms along a unified \emph{privacy--utility--attackability} trade-off curve, using the Attacker Success Rate (ASR) to quantify privacy leakage under data reconstruction attacks~\cite{Gursoy2022,Arcolezi2023,arcolezi2024revealing}, and the Mean Squared Error (MSE) to measure utility~\cite{wang2016using,tianhao2017,Cormode2021}. 
The rationale for choosing these metrics is further discussed in Section~\ref{sub:asr_mse_framework}.
Beyond reconstruction attacks, we further instantiate our framework to capture complementary adversarial objectives such as integrity threats through data poisoning~\cite{cao2021data}, alongside ASR and MSE (see Section~\ref{sub:results_poisoning}).

To operationalize this framework, we conduct an extensive analytical and empirical evaluation that provides new insights into the strengths and weaknesses of state-of-the-art protocols, and further enables a principled refinement of their design.
In particular, we show how to tune protocol parameters through multi-objective optimization, yielding adaptive variants that better navigate the \emph{privacy-utility-attackability} trade-off while preserving $\varepsilon$-LDP guarantees. 
Additionally, we derive the expected data reconstruction attack (\ie, analytical ASR) for three additional LDP protocols (SHE, THE and a generic Unary Encoding protocol), extending prior analyses~\cite{Gursoy2022,arcolezi2024revealing}. 
The main contributions of this paper are as follows:

\begin{itemize}
    \item We propose a \emph{novel evaluation framework} that systematically analyzes LDP protocols under the joint lens of adversarial success (targeting privacy leakage or integrity) and utility loss. 
    Our benchmark is designed to assist practitioners in deploying LDP protocols with customized \emph{privacy-utility-attackability} trade-offs, acting as a tuning layer adaptable to various systems (\eg, telemetry~\cite{microsoft}).
    
    \item We derive \emph{new analytical closed-form equation} of the expected data reconstruction attack for three LDP protocols, going beyond previous works~\cite{Gursoy2022}. 
    These derivations enable principled reasoning about adversarial leakage and support optimization at the protocol-design level.
    
    \item We conduct an \emph{extensive experimental evaluation} of eight state-of-the-art LDP frequency estimation protocols, quantifying their behavior across a wide range of domain sizes and privacy regimes. 
    Our results expose previously underexplored vulnerabilities and trade-offs.

    \item Leveraging the insights from our analytical and empirical analyses, we propose \emph{refined adaptive variants} (ASS, AUE, ALH, ATHE) of four well-known protocols (SS, OUE, OLH, THE). 
    Our adaptive versions consistently achieve more favorable \emph{privacy-utility-attackability} trade-offs, substantially reducing adversarial success rates while maintaining competitive estimation accuracy.
\end{itemize}

The code used in our experiments is available in our \textbf{GitHub repository}: \url{https://github.com/hharcolezi/LDP_protocols_refined}.

\section{Related Work} 
\label{sec:rel_work}

Frequency estimation is one of the primary objective of LDP that supports many downstream applications such as heavy hitter estimation~\cite{Bassily2015,Li2024,Zhang2025}, frequency monitoring~\cite{rappor,microsoft,Arcolezi2022,Arcolezi2023evolving}, multidimensional queries~\cite{Gu2019,yang2020answering,Arcolezi_rs_fd,Filho2023}, frequent item-set mining~\cite{Li2022,Wu2023}, and spatial density estimation~\cite{alptekin2023building,Tire2024}.
A large body of work has proposed LDP mechanisms to improve estimation accuracy, reduce communication cost, or enable scalable aggregation~\cite{kairouz2016discrete,wang2016mutual,tianhao2017,Min2018,Hadamard,Cormode2021,Feldman2022,arcolezi2025private}. 
While these studies have established strong theoretical and practical baselines, they primarily evaluate protocols through single-objective metrics such as MSE or communication overhead.

More recently, several works have begun to examine the adversarial dimension of LDP, including data reconstruction attacks~\cite{Gursoy2022,Arcolezi2023,arcolezi2024revealing}, re-identification risks~\cite{Murakami2021,Arcolezi2023}, attribute inference~\cite{Gadotti2022,Arcolezi2023evolving,GURSOY2024}, and poisoning attacks~\cite{cao2021data}. 
These studies revealed significant vulnerabilities in existing protocols but largely stopped at qualitative observations or empirical demonstrations, without providing a unified benchmark to compare privacy leakage and utility across mechanisms.

Our work differs by taking an \emph{experimental and analytical benchmarking} perspective. 
We quantify vulnerabilities through privacy and integrity threats and analyze their interplay with MSE across eight representative protocols. 
This joint evaluation highlights protocol-dependent trade-offs for \emph{privacy-utility-attackability} and offers a reproducible methodology for attack-aware benchmarking. 
Unlike prior analyses~\cite{Gursoy2022}, which derive ASR for a subset of mechanisms, our study extends the analytical treatment to additional protocols and demonstrates how these insights can guide the parameter tuning of existing mechanisms, yielding refined adaptive variants.

\section{Background} 
\label{sec:background}

\noindent\textbf{Notation.} We use italic uppercase letters (\eg, $U$) to denote sets, and write $[n] = \{1, \ldots, n\}$ to represent a set of $n$ positive integers. 
Vectors are denoted by bold lowercase letters (\eg, $\mathbf{x}$), where $\mathbf{x}_i$ represents the value of the $i$-th coordinate of $\mathbf{x}$. 
Finally, randomized mechanisms are denoted by $\calM$, the input domain is denoted by $\calX$, and the output domain by $\calY$. 
Both $\calX$ and $\calY$ are discrete, in which $|\calX| = k$ and $|\calY|$ depends on the randomized mechanism $\calM$.

\subsection{Problem Statement} \label{sub:problem_statement}

We consider an untrusted server collecting data from a distributed group of users while preserving their privacy. 
Formally, there are $n$ users, with each user holding a value $x$ from a discrete domain $\calX = \{1, \ldots, k\}$. 
The task is \textbf{frequency estimation}, in which the server aims to learn the frequencies of each value across all users, denoted as $\bff = \{f_i\}_{i \in [k]}$.

\begin{itemize}
    \item \textbf{Users' goal.} Each user wants to protect their privacy. 
    To achieve this, users apply an obfuscation mechanism $\calM$ that perturbs their value $x$ before sending it to the server. 
    
    \item \textbf{Server's objective.} The server aims to estimate the frequency distribution $\bff$ of the values held by all users while minimizing the estimation error. 
    After receiving the obfuscated values from all $n$ users, the server estimates a $k$-bins histogram $\hat{\bff} = \{\hat{f}_i\}_{i \in [k]}$, representing the estimated frequencies.
    
    \item \textbf{Adversary's goal.} The adversary seeks to exploit the obfuscated reports to break the system guarantees, either by inferring users’ true values (\ie, data reconstruction) or by manipulating the aggregation process to bias the estimated frequencies (\ie, data poisoning).

\end{itemize}

This threat model captures a worst-case adversary (\eg, a curious or compromised server or a man-in-the-middle) who observes or injects obfuscated reports in an untrusted setting.

\subsection{Local Differential Privacy} \label{sub:ldp}

Local Differential Privacy (LDP)~\cite{first_ldp} ensures that the output of a randomized mechanism does not significantly reveal information about the input.
Formally:

\begin{definition}[$\varepsilon$-Local Differential Privacy]
An algorithm $\calM$ satisfies $\varepsilon$-local differential privacy ($\varepsilon$-LDP), where $\varepsilon \geq 0$, if and only if for any two distinct inputs $x, x' \in \calX$, we have:
\begin{equation} \label{eq:ldp}
    \forall y \in \text{Range}(\calM) : \Pr[\calM(x) = y] \leq \eeps \cdot \Pr[\calM(x') = y] \mathrm{,}
\end{equation}
\noindent in which $\text{Range}(\calM)$ denotes the set of all possible outputs of $\calM$.
\end{definition}
Smaller values of $\varepsilon$ indicate stronger privacy guarantees, as they limit how much more likely the output is for one input compared to another given the same observed value. 
In other words, $\varepsilon$ controls the level of indistinguishability between inputs $x$ and $x'$, providing a formal measure of privacy.

\subsection{Pure LDP Framework} \label{sub:pure_ldp}

We consider the pure LDP framework proposed by Wang and co-authors~\cite{tianhao2017} to analyze LDP frequency estimation protocols. 
Formally, an LDP protocol is called pure if it satisfies the following definition:

\begin{definition}[Pure LDP Protocols~\cite{tianhao2017}]
A protocol is considered \textit{pure} if and only if there exist two probability values, $\pstar > \qstar$, in its perturbation mechanism $\calM$, such that for all inputs $x \in \calX$:
\begin{equation*}
    \begin{aligned}
        \Pr[\calM(x) \in \{y \mid x \in \text{Support}(y)\}] & = \pstar\\
        \forall_{x' \neq x} \Pr[\calM(x') \in \{y \mid x \in \text{Support}(y)\}] & = \qstar
    \end{aligned}
\end{equation*}

\noindent in which the set $\{y \mid x \in \text{Support}(y)\}$ includes all possible outputs $y$ that ``support'' the input value $x$.
\end{definition}
Thus, $\pstar$ represents the probability that the input $x$ is mapped to an output supporting it, whereas $\qstar$ represents the probability that any other input $x' \neq x$ is mapped to an output that supports $x$. 
Given $n$ users, let $y^j$ denote the obfuscated value of user $j \in [n]$, the frequency estimate $\hat{f}_i$ of the input value $i \in [k]$ is computed as:
\begin{equation} \label{eq:estimator}
    \hat{f}_i = \frac{\sum_{j=1}^{n} \setOne_{\text{Support}(y^{j})}(i) - n \qstar}{n (\pstar - \qstar)} \mathrm{.}
\end{equation}

As shown in~\cite{tianhao2017}, the estimator in Eq.~\eqref{eq:estimator} is unbiased (\ie, $\mathbb{E}[\hat{f}_i]=f_i$) and the variance of the estimation $\hat{f}_i$ is:
\begin{equation} \label{eq:var}
    \mathrm{Var}[\hat{f}_i] = \frac{\qstar (1 - \qstar)}{n (\pstar - \qstar)^2} + \frac{f_i (1 - \pstar - \qstar)}{n (\pstar - \qstar)} \mathrm{.}
\end{equation}

With a sufficiently large domain size and no dominant frequency $f_i$, the second term of Eq.~\eqref{eq:var} can be ignored. 
Thus, as commonly used in the LDP literature~\cite{tianhao2017,Arcolezi2022,yang2020answering}, we will consider the approximate variance, which is given by:
\begin{equation} \label{eq:approx_var}
    \mathrm{Var}[\hat{f}_i] = \frac{\qstar (1 - \qstar)}{n (\pstar - \qstar)^2} \mathrm{.}
\end{equation}

Furthermore, as the estimation is unbiased, we will interchangeably refer to the variance as the MSE:
\begin{equation*} \label{eq:mse_var}
  \mathrm{MSE}
  = \frac{1}{k} \sum_{i=1}^{k} \mathbb{E} \left[ \left( \hat{\bff}_i - \bff_i \right)^2 \right ]
  = \frac{1}{k} \sum_{i=1}^{k} \mathrm{Var}[\hat{\bff}_i] \mathrm{.}
\end{equation*}

\subsection{Data Reconstruction Attack on LDP} \label{sub:bayes_attack_ldp}

The fundamental premise of $\varepsilon$-LDP is that the input to $\calM$ cannot be confidently inferred from its output, with the level of confidence bounded by $\eeps$. 
A user's privacy is thus compromised if an adversary $\calA$ can accurately infer the original input $x \in \calX$ from the obfuscated output $y \in \calY$.
To assess this residual inference risk, we focus on \emph{data reconstruction attacks}~\cite{Gursoy2022,Arcolezi2023,arcolezi2024revealing}, which directly quantify how well an adversary can recover the true input from the perturbed report.

Formally, the adversary infers: 
\begin{equation}
\hat{x} = \arg\max_{x \in \calX} \Pr[x \mid y].
\end{equation}

Applying Bayes' theorem gives

\begin{equation} \label{eq:bayes_attack_general_prior}
\hat{x} = \arg\max_{x \in \calX} \Pr[y \mid x] \cdot \Pr[x] \mathrm{.}
\end{equation}

Without loss of generality, we adopt a \textbf{uniform prior}, \ie, $\Pr[x] = 1/k$ with $k = |\mathcal{X}|$, to obtain a protocol-intrinsic and distribution-agnostic measure of attackability.
Thus, Eq.~\eqref{eq:bayes_attack_general_prior} reduces to
\begin{equation} \label{eq:bayes_attack}
\hat{x} = \arg\max_{x \in \mathcal{X}} \Pr[y \mid x],
\end{equation}
so that the adversary’s decision depends solely on the mechanism’s likelihood function.
As shown in prior work~\cite{Gursoy2022}, incorporating informative priors naturally increases the attack success rate by injecting side information, motivating our conservative distribution-independent choice.

We quantify reconstruction risk using the \textbf{Attacker Success Rate (ASR)}, defined as the probability that an adversary correctly reconstructs a user’s original input~\cite{Gursoy2022}:
\begin{equation} \label{eq:ASR}
\text{ASR} = \Pr[\hat{x} = x] = \frac{1}{n} \sum_{j=1}^{n}\setOne(\hat{x}^j = x^j) \mathrm{,}
\end{equation}
\noindent which offers a more interpretable view of privacy protection, revealing vulnerabilities not captured by $\varepsilon$.

\textbf{Connection to other inference attacks.}
Data reconstruction constitutes a \emph{fundamental inference task} under LDP.
Other attacks, such as attribute inference~\cite{Gadotti2022,Arcolezi2023evolving,GURSOY2024} and re-identification~\cite{Murakami2021,Arcolezi2023}, can be viewed as special cases that build upon reconstruction by leveraging correlations across attributes or external knowledge.
This is similar to prior formulations~\cite{Balle2022,Guerra2024} in the machine learning community relating data reconstruction to attribute and membership inference.

\begin{example}[Reconstruction as a proxy for re-identification] \label{ex:re_identification}
Consider a user described by $d$ quasi-identifiers $x = (x^{(1)}, \dots, x^{(d)})$, each independently perturbed under an $\varepsilon$-LDP mechanism.
Let $\hat{x}^{(i)}$ denote the adversary’s reconstruction of attribute $x^{(i)}$ from the corresponding privatized report.
The joint reconstruction success is defined as~\cite{Arcolezi2023}:
\[
\mathrm{ASR}_{\mathrm{joint}} = \Pr[\hat{x}^{(1)} = x^{(1)} \wedge \ldots \wedge \hat{x}^{(d)} = x^{(d)}].
\]
Under the conditional independence of the reconstructions, this reduces to
$\mathrm{ASR}_{\mathrm{joint}} = \prod_{i=1}^{d} \mathrm{ASR}_i$.
Correctly reconstructing all quasi-identifiers might uniquely characterize a record within a population~\cite{Samarati1998}, making $\mathrm{ASR}_{\mathrm{joint}}$ a natural proxy for linkage attacks and re-identification.
\end{example}

\begin{example}[Reconstruction as a proxy for attribute inference] \label{ex:attribute_inference}
Consider an iterative data collection setting in which a user reports their value $x_t \in \mathcal{X}$ at each time step $t$ using an $\varepsilon$-LDP mechanism.
Following standard deployments based on \emph{memoization} (\eg, Microsoft's $d$BitFlipPM~\cite{microsoft}), the same privatized output is reused across time as long as the underlying value remains unchanged.
Let $\hat{x}_t$ denote the adversary’s reconstruction at time $t$.
If reconstruction succeeds with non-negligible probability, the adversary can infer data changes by detecting transitions in the reconstructed outputs~\cite{Arcolezi2023evolving}:
\[
\Pr[\hat{x}_t \neq \hat{x}_{t-1}] \approx \Pr[x_t \neq x_{t-1}].
\]

As a result, successful reconstruction enables attribute inference over time, even when individual reports satisfy $\varepsilon$-LDP~\cite{GURSOY2024,Gadotti2022}.
This allows an adversary to infer sensitive behavioral changes (\eg, starting a diet or going on holiday) without necessarily recovering the exact value itself.
\end{example}

\section{Overview of Pure LDP Frequency\\Estimation Protocols} \label{sec:ldp_protocols}
In this section, we provide a concise overview of eight state-of-the-art pure LDP frequency estimation protocols, selected as representative mechanisms underlying widely deployed industrial systems (\eg, SUE-based in Microsoft~\cite{microsoft} telemetry, SS in Google Gboard~\cite{sun2024private} and OLH-based in Apple~\cite{apple}).
We systematically describe each mechanism based on three key functions: \textbf{Encoding}, \textbf{Perturbation} and \textbf{Aggregation}, which collectively define their operation. 
Additionally, we introduce an \textbf{Attacking} function for each protocol, highlighting potential vulnerabilities to privacy attacks.
For protocols in which the expected ASR is not available in the literature~\cite{Gursoy2022}, we derive it and present the detailed analyses in Appendix~\ref{app:expected_asr}.

\subsection{Generalized Randomized Response (GRR)} \label{sub:grr_protocol}
GRR~\cite{kairouz2016discrete} extends the randomized response method proposed by Warner~\cite{Warner1965} to a domain size of $k \geq 2$, while ensuring $\varepsilon$-LDP.

\textbf{Encoding.} In GRR, $\texttt{Encode}(x)=x$ and $x \in [k]$.

\textbf{Perturbation.} The perturbation function of GRR is:
\begin{equation} \label{eq:grr}
    \Pr[\mathrm{GRR}(x)=y] = \begin{cases} p=\frac{\eeps}{\eeps+k-1} \textrm{ if } y = x,\\ q=\frac{1}{\eeps+k-1} \textrm{ if } y \neq x \textrm{,} \end{cases}
\end{equation}
\noindent in which $y \in [k]$ is the perturbed value sent to the server.

\textbf{Aggregation.} In GRR, each output value $i$ supports the corresponding input $i$, resulting in the support set $\setOne_{\mathrm{GRR}} = {y}$. 
GRR is a pure protocol with $\pstar = p$ and $\qstar = q$. 
The server estimates the frequency using Eq.~\eqref{eq:estimator}, with the following analytical MSE:
\begin{equation} \label{eq:var_grr}
    \mathrm{MSE}_{\mathrm{GRR}} = \frac{\eeps + k - 2}{n(\eeps - 1)^2} \mathrm{.}
\end{equation}

\textbf{Attacking.} From Eq.~\eqref{eq:grr}, it follows that $\Pr[y = x] > \Pr[y = x']$ for all $x' \in \calX \setminus \{x\}$. 
Thus, the optimal attack strategy for GRR is to predict $\hat{x} = \calA_{\mathrm{GRR}}(y) = y$. 
The expected ASR for GRR is given by:
\begin{equation} \label{eq:exp_asr_grr}
    \text{ASR}_{\mathrm{GRR}} = \frac{\eeps}{\eeps+k-1} \mathrm{.}
\end{equation}

\subsection{Subset Selection (SS)}  \label{sub:ss_protocol}

SS~\cite{wang2016mutual,Min2018} outputs a randomly selected subset $\mathbf{y}$ of size $\omega$ from the original domain $\calX$. 
SS can be seen as a generalization and optimization of GRR, in which SS is equivalent to GRR when $\omega = 1$.

\textbf{Encoding and Perturbation.} Starting with an empty subset $\mathbf{y} = \emptyset$, the true value $x$ is added to $\mathbf{y}$ with probability: $p=\frac{\omega \eeps}{\omega \eeps + k - \omega}$. 
Finally, values are added to $\mathbf{y}$ as follows:

\begin{itemize}
    \item If $x \in \mathbf{y}$, then $\omega - 1$ values are sampled from $\calX \setminus \{x\}$ uniformly at random (without replacement) and are added to $\mathbf{y}$;
    
    \item If $x \notin \mathbf{y}$, then $\omega$ values are sampled from $\calX \setminus \{x\}$ uniformly at random (without replacement) and are added to $\mathbf{y}$.
\end{itemize}

The user then sends the subset $\mathbf{y}$ to the server.

\textbf{Aggregation.} In SS, each value $i$ in the output subset $\mathbf{y}$ supports the corresponding input value $i$. 
Thus, the support set for SS is $\setOne_{\mathrm{SS}} = \{x \mid x \in \mathbf{y}\}$. 
This protocol is pure, with: $\pstar=p=\frac{\omega e^{\varepsilon}}{\omega e^{\varepsilon} + k - \omega}$ and $\qstar=\frac{\omega e^{\varepsilon} (\omega - 1) + (k - \omega) \omega}{(k - 1) (\omega e^{\varepsilon} + k - \omega)}$.
The server estimates the frequency using Eq.~\eqref{eq:estimator}, with the following analytical MSE:
\begin{equation} \label{eq:var_ss}
\begin{aligned}
    \mathrm{MSE}_{\mathrm{SS}} = \Bigg(
    \frac{
        (k - 1)(k + 2\omega e^{\varepsilon} - \omega)
    }{
        n \omega \left( -k + \omega + (k - 1) e^{\varepsilon} - (\omega - 1) e^{\varepsilon} \right)^2
    } \\
    \left. + \frac{
        \left(k - \omega + (\omega - 1) e^{\varepsilon}\right)
        \left(
            - \omega (k - \omega) - \omega (\omega - 1) e^{\varepsilon} \right)
    }{
        n \omega \left( -k + \omega + (k - 1) e^{\varepsilon} - (\omega - 1) e^{\varepsilon} \right)^2
    }
        \right) \mathrm{.}
    \end{aligned}
\end{equation}

The optimal subset size that minimizes the MSE of SS in Eq.~\eqref{eq:var_ss} is $\omega = \max \left(1, \left\lfloor \frac{k}{\eeps + 1} \right\rceil \right)$~\cite{wang2016mutual, Min2018}.

\textbf{Attacking.} With the support set of each user’s report $\setOne_{\mathrm{SS}}$, the optimal attack strategy $\calA_{\mathrm{SS}}$ is to predict $\hat{x}=\mathrm{Uniform}\left( \setOne_{\mathrm{SS}} \right)$~\cite{Gursoy2022,Arcolezi2023}.
The expected ASR for SS is~\cite{Gursoy2022}:
\begin{equation} \label{eq:exp_asr_ss}
    \text{ASR}_{\mathrm{SS}} = \frac{e^{\varepsilon}}{\omega e^{\varepsilon} + k - \omega} \mathrm{.}
\end{equation}

\subsection{Unary Encoding (UE)} \label{sub:ue_protocols}
UE protocols~\cite{rappor,tianhao2017} encode the user's input $x \in \calX$ as a $k$-dimensional one-hot vector $\mathbf{x}$, in which each bit is subsequently obfuscated independently.

\textbf{Encoding.} $\texttt{Encode}(x) = [0, \ldots, 0, 1, 0, \ldots, 0]$ is a binary vector with a single $1$ at position $x$ and all other positions set to $0$.

\textbf{Perturbation.} The obfuscation function of UE mechanisms randomizes the bits from $\textbf{x}$ independently to generate $\textbf{y}$ as:
\begin{equation}  \label{eq:ue_parameters}
    \forall{i \in [k]} : \quad \Pr[\textbf{y}_i=1] =\begin{cases} p, \textrm{ if } \textbf{x}_i=1 \textrm{,} \\ q, \textrm{ if } \textbf{x}_i=0 \textrm{,}\end{cases}
\end{equation}
\noindent in which $\textbf{y}$ is sent to the server. 
There are two variations of UE mechanisms: (i) Symmetric UE (SUE)~\cite{rappor} that selects $p=\frac{e^{\varepsilon/2}}{e^{\varepsilon/2}+1}$ and $q=\frac{1}{e^{\varepsilon/2}+1}$; and (ii) Optimized UE (OUE)~\cite{tianhao2017} that selects $p=\frac{1}{2}$ and $q=\frac{1}{\eeps+1}$ to minimize the MSE in Eq.~\eqref{eq:var_ue} below.

\textbf{Aggregation.} A reported bit vector $\mathbf{y}$ is considered to support an input $i$ if $\mathbf{y}_i = 1$.
Therefore, the support set for UE protocols is defined as $\setOne_{\mathrm{UE}}=\{i | \textbf{y}_i = 1\}$.
UE protocols are pure with $\pstar=p$ and $\qstar=q$.
The server estimates the frequency using Eq.~\eqref{eq:estimator}, with the corresponding MSE as:
\begin{equation} \label{eq:var_ue}
    \mathrm{MSE}_{\mathrm{UE}} = \frac{\left( \left( e^{\varepsilon} - 1 \right) q + 1 \right)^2}{n\left( e^{\varepsilon} - 1 \right)^2 (1 - q) q} \mathrm{.}
\end{equation}

\textbf{Attacking.} Given the support set of each user's report, $\setOne_{\mathrm{UE}}$, the adversary can adopt two possible attack strategies $\calA_{\mathrm{UE}}$~\cite{Gursoy2022, Arcolezi2023}:

\begin{itemize}
    \item $\calA^0_{\mathrm{UE}}$ is a random choice $\hat{x}=\mathrm{Uniform}\left( [k] \right)$, if $\setOne_{\mathrm{UE}}=\emptyset$;
    
    \item $\calA^1_{\mathrm{UE}}$ is a random choice $\hat{x}=\mathrm{Uniform}\left( \setOne_{\mathrm{UE}} \right)$, otherwise.
\end{itemize}

In this paper, we generalized the expected ASR of SUE and OUE given in~\cite{Gursoy2022} for any UE protocol as:
\begin{equation} \label{eq:exp_asr_ue}
\mathrm{ASR}_{\mathrm{UE}}
= \frac{1}{k}\left[(1-p)(1-q)^{k-1} + \frac{p}{q}\left(1-(1-q)^{k}\right)\right] 
\end{equation}

We defer the derivation of Eq.~\eqref{eq:exp_asr_ue} to Appendix~\ref{app:asr_ue}.

\subsection{Local Hashing (LH)}  \label{sub:lh_protocols}
LH protocols~\cite{tianhao2017,Bassily2015} use hash functions to map the input data $x \in \calX$ to a new domain $[g]$, and then obfuscate the hash value with GRR. 
Let $\mathscr{H}$ be a universal hash function family such that each hash function $\mathrm{H} \in \mathscr{H}$ hashes a value $x \in \calX$ into $[g]$ (\ie, $\mathrm{H} : [k] \rightarrow [g]$). 

\textbf{Encoding.} $\texttt{Encode}(x) = \langle \mathrm{H}, h \rangle$, in which $\mathrm{H} \in \mathscr{H}$ is chosen uniformly at random, and $h=\mathrm{H}(x)$.
There are two variations of LH mechanisms: (i) Binary LH (BLH)~\cite{Bassily2015} that just sets $g=2$, and (ii) Optimized LH (OLH)~\cite{tianhao2017} that selects $g=\lfloor \eeps + 1 \rceil$ to minimize the MSE in Eq.~\eqref{eq:var_lh} below.

\textbf{Perturbation.} LH protocols perturb $\langle \mathrm{H}, h \rangle$ into $\langle \mathrm{H}, y \rangle$, just like GRR, as follows:
\begin{equation}
    \forall i \in [g], \; \Pr[y = i] = 
    \begin{cases}
        p = \frac{e^{\varepsilon}}{e^{\varepsilon} + g - 1}, & \text{if } h = i \mathrm{,} \\
        q = \frac{1}{e^{\varepsilon} + g - 1}, & \text{if } h \neq i \mathrm{.}
    \end{cases}
\end{equation}

\textbf{Aggregation.} For each reported tuple $\langle \mathrm{H}, y \rangle$, the support set for LH protocols consists of all values $x \in \calX$ that hash to $y$, denoted as $\setOne_{\mathrm{LH}}= \{x \mid \mathrm{H}(x) = y\}$.
LH protocols are pure with $\pstar=p$ and $\qstar=\frac{1}{g}$.
The server estimates the frequency using Eq.~\eqref{eq:estimator} with the following analytical MSE:
\begin{equation} \label{eq:var_lh}
    \mathrm{MSE}_{\mathrm{LH}} = \frac{(e^{\varepsilon} - 1 + g)^2}{n(e^{\varepsilon} - 1)^2 (g - 1)} \mathrm{.}
\end{equation}

\textbf{Attacking.} Based on the support set of each user's report, $\setOne_{\mathrm{LH}}$, the adversary can employ one of two possible attack strategies, denoted by $\calA_{\mathrm{LH}}$~\cite{Gursoy2022, Arcolezi2023}:

\begin{itemize}
    \item $\calA^0_{\mathrm{LH}}$ is a random choice $\hat{x}=\mathrm{Uniform}\left( [k] \right)$, if $\setOne_{\mathrm{LH}}=\emptyset$;
    
    \item $\calA^1_{\mathrm{LH}}$ is a random choice $\hat{x}=\mathrm{Uniform}\left( \setOne_{\mathrm{LH}} \right)$, otherwise.
\end{itemize}

The expected ASR of LH protocols is given by~\cite{Gursoy2022}:
\begin{equation}  \label{eq:exp_asr_lh}
    \text{ASR}_{\mathrm{LH}} = \frac{e^{\varepsilon}}{(e^{\varepsilon} + g - 1) \cdot \max \left\{ \frac{k}{g}, 1 \right\}} \mathrm{.}
\end{equation}

\subsection{Histogram Encoding (HE)}
HE protocols~\cite{tianhao2017} encode the user's input data $x \in \calX$, as a one-hot $k$-dimensional histogram before obfuscating each bit independently.

\textbf{Encoding.} $\texttt{Encode}(x) = [0.0, 0.0, \ldots, 1.0, 0.0, \ldots, 0.0]$ in which only the $x$-th component is $1.0$.
Two different input values $x, x' \in \calX$ will result in two vectors with L1 distance of $\Delta_1=2$. 

\textbf{Perturbation.} The perturbation function $\texttt{Perturb}(\mathbf{x})$ generates the output vector $\mathbf{y}$, where each component is given by $\textbf{y}_i = \textbf{x}_i + \mathrm{Lap}\left( \frac{\Delta_1}{\varepsilon} \right)$, with $\mathrm{Lap}(\cdot)$ representing the Laplace mechanism~\cite{Dwork2006}.

The following subsections describe two HE-based mechanisms: Summation with HE (SHE) and Thresholding with HE (THE). 
These mechanisms differ in their aggregation and attack strategies.

\subsubsection{Summation with HE (SHE)} \label{sub:she_protocol}
With SHE, there is no post-processing of $\textbf{y}$ at the server side.

\textbf{Aggregation.} Since Laplace noise with mean $0$ is added to each vector independently, the server estimates the frequency using the sum of the noisy reports: $\hat{\bff} = \left\{ \sum_{j=1}^{n} \mathbf{y}^j_i \right\}_{i \in [k]}$.
This aggregation method for SHE does not provide a support set and is not pure.
The analytical MSE of this estimation is:
\begin{equation} \label{eq:var_she}
    \mathrm{MSE}_{\mathrm{SHE}}=\frac{8}{n\varepsilon^2} \mathrm{.}
\end{equation}

\textbf{Attacking.} The optimal attack strategy for SHE is to predict the user's value by selecting the index corresponding to the maximum component of the obfuscated vector: $\hat{x} = \underset{i \in [k]}{\mathrm{argmax}} \; y_i$~\cite{arcolezi2024revealing}.
Following this attack strategy, we deduce the expected ASR for the SHE protocol as the probability that the noisy value \( y_x \) at the true index exceeds all other noisy values \( y_i \) for \( i \neq x \):
\begin{equation} \label{eq:exp_asr_she}
\text{ASR}_{\mathrm{SHE}} = \Pr \left[ y_x > \max_{i \neq x} y_i \right].
\end{equation}

We defer the derivation of Eq.~\eqref{eq:exp_asr_she} to Appendix~\ref{app:asr_she}.

\subsubsection{Thresholding with HE (THE)} \label{sub:the_protocol}

In THE, each perturbed component of the vector $\mathbf{y}$ is compared to a threshold value $\theta$ to generate the final output vector.
More precisely:
\begin{equation*}
\forall{i \in [k]} : \quad \mathbf{y}_i = \begin{cases}
1, & \text{if } \mathbf{y}_i > \theta \\
0, & \text{if } \mathbf{y}_i \leq \theta
\end{cases}
\end{equation*}

Thus, the resulting output vector $\mathbf{y}$ is a binary vector in $\{0, 1\}^k$, where we have the following probabilities:
\begin{align} \label{eq:the_parameters}
    p &= \Pr[\mathbf{y}_i = 1 \mid \mathbf{x}_i = 1] = 1 - \frac{1}{2} e^{\frac{\varepsilon}{2} (\theta - 1)} \mathrm{.} \\
    q &= \Pr[\mathbf{y}_i = 1 \mid \mathbf{x}_i = 0] = \frac{1}{2} e^{-\frac{\varepsilon}{2} \theta} \mathrm{.} 
\end{align}

\textbf{Aggregation.} A reported bit vector $\mathbf{y}$ is viewed as supporting an input $i$ if $\mathbf{y}_i > \theta$.
Therefore, the support set for THE is $\setOne_{\mathrm{THE}} = \{ i \hspace{0.1cm} | \hspace{0.1cm} \textbf{y}_i \hspace{0.1cm} >  \hspace{0.1cm}\theta\}$.
The THE mechanism is pure with $\pstar=p$ and $\qstar=q$.
The server estimates the frequency using Eq.~\eqref{eq:estimator} with the following analytical MSE:
\begin{equation} \label{eq:var_the}
    \mathrm{MSE}_{\mathrm{THE}} = \frac{2 e^{\varepsilon \theta / 2} - 1}{n(1 + e^{\varepsilon (\theta - 1/2)} - 2 e^{\varepsilon \theta / 2})^2} \mathrm{.}
\end{equation}

The optimal threshold value that minimizes the protocol's MSE in Eq.~\eqref{eq:var_the} is within $\theta \in (0.5, 1)$~\cite{tianhao2017}.

\textbf{Attacking.} Based on the support set $\setOne_{\mathrm{THE}}$, the adversary can use one of two attack strategies denoted by $\calA_{\mathrm{THE}}$~\cite{arcolezi2024revealing}:

\begin{itemize}
    \item $\calA^0_{\mathrm{THE}}$ is a random choice $\hat{x}=\mathrm{Uniform}\left( [k] \right)$, if $\setOne_{\mathrm{THE}}=\emptyset$;
    
    \item $\calA^1_{\mathrm{THE}}$ is a random choice $\hat{x}=\mathrm{Uniform}\left( \setOne_{\mathrm{THE}} \right)$, otherwise.
\end{itemize}

In this paper, we obtained the expected ASR for THE as:
\begin{equation} \label{eq:exp_asr_the}
    \mathrm{ASR}_{\mathrm{THE}} = (1 - p)(1-q)^{k-1} \cdot \frac{1}{k} + \frac{p}{kq}\left(1 - (1-q)^k\right) \mathrm{.}
\end{equation}

We defer the derivation of Eq.~\eqref{eq:exp_asr_the} to Appendix~\ref{app:asr_the}.

\section{Attack-Aware Evaluation Framework for LDP} \label{sec:our_protocols}

The existing literature has traditionally proposed mechanisms like SS, OUE, OLH and THE, with the primary objective of minimizing estimation error for a given privacy budget $\varepsilon$. 
However, protocols offering identical $\varepsilon$-LDP guarantees can exhibit substantially different vulnerabilities to inference and poisoning attacks, leading to markedly different real-world privacy and integrity risks. Motivated by this observation, we propose an \emph{attack-aware evaluation framework} for LDP frequency estimation protocols, designed to systematically analyze how estimation utility interacts with adversarial vulnerability under different threat models and protocol configurations.

\subsection{General Evaluation Framework} \label{sub:multi_objective_framework}

We evaluate LDP protocols through a unified \emph{attack-aware} instantiation, in which each mechanism is characterized by its behavior along two complementary axes beyond $\varepsilon$, capturing key system desiderata:
\begin{equation} \label{eq:multi_objective_framework}
\left( \mathcal{A}(\theta), \; \mathcal{U}(\theta) \right),
\end{equation}
in which $\theta \in \Theta$ denotes the protocol parameters (\eg, $\varepsilon$ or design choices), $\mathcal{U}$ measures utility loss and $\mathcal{A}$ quantifies the protocol \emph{attackability} under a specified adversarial model.
Specifically, the attackability component $\mathcal{A}$ can be instantiated to capture different classes of threats:
\begin{itemize}
    \item \textbf{Privacy attackability} ($\mathcal{A}_{\mathrm{priv}}$), including data reconstruction~\cite{Gursoy2022,arcolezi2024revealing}, attribute inference~\cite{Gadotti2022,Arcolezi2023evolving,GURSOY2024} or re-identification~\cite{Arcolezi2023,Murakami2021};
    
    \item \textbf{Integrity attackability} ($\mathcal{A}_{\mathrm{sec}}$), such as vulnerability to data poisoning~\cite{cao2021data} or adversarial manipulation~\cite{Cheu2021}.
\end{itemize}

Utility loss $\mathcal{U}$ is typically quantified using metrics based on estimation error, such as MSE, Mean Absolute Error (MAE)~\cite{kairouz2016discrete}, or Fisher Information~\cite{Barnes2020}.

This formulation provides a principled lens for reasoning about the inherent trade-offs between estimation accuracy and adversarial vulnerability in the design of LDP protocols.

\subsection{Specialization to Data Reconstruction and MSE Trade-off} \label{sub:asr_mse_framework}

In this work, we focus on a widely applicable and analytically tractable instantiation that captures the fundamental privacy-utility tension in LDP.
Specifically, we consider:

\begin{itemize}
    \item \textbf{Privacy attackability}, measured via the ASR under \emph{data reconstruction attacks} in Eq.~\eqref{eq:ASR}.
    Reconstruction attacks directly target users' private values and admit closed-form analyses for all eight state-of-the-art LDP protocols considered in this paper.
    Moreover, successful reconstruction can serve as a proxy for other downstream privacy risks~\cite{Balle2022,Guerra2024}, such as profiling and re-identification (see Example~\ref{ex:re_identification}) as well as attribute inference (see Example~\ref{ex:attribute_inference}), making ASR a natural and interpretable indicator of broader privacy leakage.
    
    \item \textbf{Utility}, measured by the variance (MSE) of frequency estimation, which directly quantifies estimation accuracy and is central to frequency estimation tasks~\cite{wang2016using,tianhao2017,Cormode2021}.
    Minimizing MSE also implicitly controls other utility losses such as MAE and Fisher Information through established theoretical relationships like the Central Limit Theorem and Cramér-Rao bounds~\cite{rao1992information,CRAMÉR1999,Cover2006}. 
\end{itemize}

This ASR-MSE instantiation yields a meaningful privacy-utility frontier that is central to many practical LDP deployments, and leverages closed-form expressions available for a wide range of protocols (see Section~\ref{sec:ldp_protocols}).

\textbf{Protocol Selection via Pareto Optimization.}
Since ASR and MSE represent inherently competing objectives, each LDP protocol can be characterized by a set of \textbf{Pareto-efficient configurations} that balance privacy leakage and utility. 
To identify these configurations, we evaluate both metrics across a predefined parameter grid~\cite{bergstra2012random}, forming the analytical Pareto frontier that captures the achievable trade-offs for each mechanism. 
Once the Pareto frontier is obtained, we select representative operating points using established multi-objective selection criteria (\eg, Utopia Point, Weighted Scalarization, Elbow ``Knee'' Method, etc)~\cite{Donoso2007}.
Among these, \textbf{weighted scalarization} stands out as a practical default due to its conceptual simplicity, interpretability and tunable nature.
Empirically, we observed that all selection methods, except for the Elbow and $\boldsymbol{\epsilon}$-Constraint methods, tend to identify similar parameters (see Figure~\ref{fig:appendix_optimization} in Appendix~\ref{app:optimization_methods}).  
For clarity and consistency, we therefore report results using the scalarized formulation:
\begin{equation} \label{eq:objective_function}
    \begin{aligned}
        & \underset{\mathbf{\Theta}}{\mathrm{min}}
        & & J(\mathbf{\Theta}) = w_{\text{ASR}} \cdot \text{ASR} 
        + w_{\mathrm{MSE}} \cdot \mathrm{MSE} \mathrm{,} \\
        & \emph{s.t.}
        & & w_{\text{ASR}} + w_{\mathrm{MSE}} = 1 \mathrm{.}
    \end{aligned}
\end{equation}

Varying the weights $(w_{\text{ASR}}, w_{\mathrm{MSE}})$ enables controlled exploration of privacy-utility preferences: higher $w_{\text{ASR}}$ emphasizes stronger protection, while larger $w_{\mathrm{MSE}}$ favors utility. 
This parameterization provides a unified way to visualize trade-offs across protocols and privacy regimes. 
Importantly, setting $w_{\text{MSE}} = 1$ recovers the original protocols, showing that \emph{\textbf{our refined mechanisms generalize rather than replace}} their classical counterparts.

\begin{remark}[Attack-Agnostic Nature of the Framework]
\label{rem:attack_agnostic}

The proposed attack-aware framework in Section~\ref{sub:multi_objective_framework} is not restricted to reconstruction attacks.
More precisely, it is \emph{attack-agnostic} and can accommodate arbitrary adversarial models, provided that the corresponding attackability metric can be evaluated, either analytically or via empirical (Monte Carlo) estimation.

In Section~\ref{sub:results_poisoning}, we illustrate this flexibility by incorporating integrity-oriented attacks through data poisoning attacks~\cite{cao2021data}.
While different attack metrics may interact with protocol parameters in distinct ways, the framework naturally supports their joint analysis, either via complementary two-dimensional trade-offs or through higher-dimensional visualizations and constrained selection strategies.
\end{remark}

\subsection{Refining LDP Protocols}  \label{sub:our_protocols}
Using our two-objective framework defined in Eq.~\eqref{eq:objective_function}, we extend four state-of-the-art LDP protocols to introduce adaptive counterparts.  
Each adaptive protocol selects an optimal parameter \(\mathbf{\Theta}\) to achieve a better trade-off between ASR and MSE, rather than focusing solely on utility.  
The optimization process varies across protocols, adapting their internal parameters to balance privacy protection and estimation accuracy. 
\textit{\textbf{Importantly, our reparametrization preserves the original $\varepsilon$-LDP guarantee by design.}}
For instance, the subset size $\omega$ in ASS and the probability pair $(p, q)$ in AUE are explicitly constrained to satisfy the $\varepsilon$-LDP guarantee; 
ALH inherits the privacy guarantee of GRR applied to a hashed domain; 
and ATHE applies a post-processing threshold, which does not affect the privacy guarantee.
We now describe each adaptive mechanism in detail.  

\subsubsection{\textbf{Adaptive Subset Selection (ASS)}} 
\label{sub:ass_protocol}
The ASS mechanism extends the SS protocol.  
Unlike SS, which aims to minimize the estimation error alone when selecting $\omega$, ASS jointly optimizes it for both MSE and ASR.  
Formally, the optimization problem for ASS is defined as:
\begin{equation} \label{eq:optimization_ass}
\resizebox{\linewidth}{!}{$
\begin{aligned}
    & \underset{\omega \in \mathbb{Z}}{\mathrm{min}}
    & & J(\omega) = w_{\mathrm{MSE}} \cdot \Bigg(
    \frac{
        (k - 1)(k + 2\omega e^{\varepsilon} - \omega)
    }{
        n \omega \left( -k + \omega + (k - 1) e^{\varepsilon} - (\omega - 1) e^{\varepsilon} \right)^2
    } \\
    & & & \left. + \frac{
        \left(k - \omega + (\omega - 1) e^{\varepsilon}\right)
        \left(
            - \omega (k - \omega) - \omega (\omega - 1) e^{\varepsilon} \right)
    }{
        n \omega \left( -k + \omega + (k - 1) e^{\varepsilon} - (\omega - 1) e^{\varepsilon} \right)^2
    }
        \right)  \\
    & & & + w_{\text{ASR}} \cdot \left( \frac{e^{\varepsilon}}{\omega e^{\varepsilon} + k - \omega} \right), \\
    & \emph{s.t.}
    & & 1 \leq \omega < k, \quad w_{\text{ASR}} + w_{\mathrm{MSE}} = 1 \mathrm{.}
\end{aligned}
$}
\end{equation}

The range \( 1 \leq \omega < k \) ensures that at least one value is selected while keeping the subset smaller than the total domain size.  
This prevents trivial cases where \(\omega = k\) would result in full randomness in the response or $\omega=0$ (no report).

\subsubsection{\textbf{Adaptive Unary Encoding (AUE)}} \label{sub:aue_protocol}
The AUE mechanism is a generalization of UE protocols. 
Unlike the optimized UE protocol (\ie, OUE~\cite{tianhao2017}), which only aims to minimize the estimation error setting a fixed $p=1/2$ and $q=\frac{1}{e^{\varepsilon}+1}$, AUE jointly optimizes both MSE and ASR by adapting the probabilities \( p \) and \( q \).
Formally, the optimization problem for AUE is defined as:
\begin{equation} \label{eq:optimization_aue}
\begin{aligned}
    & \underset{p \in \mathbb{R}}{\mathrm{min}}
    & & J(p) = w_{\text{MSE}} \cdot \left( \frac{\left( \left( e^{\varepsilon} - 1 \right) q + 1 \right)^2}
    {n \left( e^{\varepsilon} - 1 \right)^2 (1 - q) q} \right)\\
    & & & + w_{\text{ASR}} \cdot \frac{1}{k}\left[(1-p)(1-q)^{k-1} + \frac{p}{q}\left(1-(1-q)^{k}\right)\right]
    \mathrm{,} \\
    & \emph{s.t.}
    & & 0.5 \leq p < 1, \;  q = \frac{p}{e^{\varepsilon} (1 - p) + p}, \; w_{\text{ASR}} + w_{\mathrm{MSE}} = 1 \mathrm{.} 
\end{aligned}
\end{equation}

The equality constraint for $q$ is due to the $\varepsilon$-LDP requirement for UE protocols: $\varepsilon= \ln \left( \frac{p (1 - q)}{(1 - p) q} \right)$~\cite{rappor,tianhao2017}.

\subsubsection{\textbf{Adaptive LH (ALH)}} \label{sub:alh_protocol}
The ALH mechanism extends the LH protocol. 
Unlike the optimized LH protocol (\ie, OLH~\cite{tianhao2017}), which aims solely to minimize estimation error by selecting $g=\lfloor \eeps + 1 \rceil$, ALH jointly optimizes both MSE and ASR by adapting the hash domain size parameter \( g \). 
Formally, the optimization problem for ALH is defined as:
\begin{equation} \label{eq:optimization_alh}
\begin{aligned}
    & \underset{g \in \mathbb{Z}}{\mathrm{min}}
    & & J(g) = w_{\text{MSE}} \cdot 
    \left( 
    \frac{(e^{\varepsilon} - 1 + g)^2}{n(e^{\varepsilon} - 1)^2 (g - 1)}
    \right) \\
    & & & \quad + w_{\text{ASR}} \cdot 
    \left( 
    \frac{e^{\varepsilon}}{(e^{\varepsilon} + g - 1) \cdot \max \left\{ \frac{k}{g}, 1 \right\}}
    \right) \mathrm{.} \\
    & \emph{s.t.}
    & & 2 \leq g \leq \max \left( k, \lfloor e^{\varepsilon} + 1 \rceil \right), \; w_{\text{ASR}} + w_{\mathrm{MSE}} = 1  \mathrm{,} 
\end{aligned}
\end{equation}

The upper bound for \( g \) is set to \( \max \left ( k, \lfloor \eeps + 1 \rceil \right) \), ensuring that \( g \) is not unnecessarily smaller than the original domain size \( k \). 
This avoids under-hashing, allowing for better differentiation and randomness. 
When \( k \) is large, \( g \) should ideally match or exceed \( k \), ensuring that hashing effectively introduces randomness to maintain privacy guarantees.

\subsubsection{\textbf{Adaptive THE (ATHE)}} \label{sub:athe_protocol}
The ATHE mechanism extends the THE protocol. 
Unlike THE~\cite{tianhao2017}, which only aims to minimize the MSE in Eq.~\eqref{eq:var_the}, ATHE jointly optimizes both MSE and ASR by adapting the threshold parameter \(\theta\). 
Formally, the optimization problem for ATHE is defined as:
\begin{equation} \label{eq:optimization_athe}
\begin{aligned}
    & \underset{\theta \in \mathbb{R}}{\mathrm{min}}
    & & J(\theta) = w_{\mathrm{MSE}} \cdot 
    \left( 
    \frac{2 e^{\varepsilon \theta / 2} - 1}{n(1 + e^{\varepsilon (\theta - 1/2)} - 2 e^{\varepsilon \theta / 2})^2}
    \right) \\
    & & & \quad + w_{\text{ASR}} \cdot \Bigg( (1 - p)(1-q)^{k-1} \cdot \frac{1}{k} + \frac{p}{kq}\left(1 - (1-q)^k\right)  \Bigg) \mathrm{,}\\
    & \emph{s.t.}
    & & 0.5 \leq \theta \leq 1, \; w_{\text{ASR}} + w_{\mathrm{MSE}} = 1  \mathrm{,}
\end{aligned}
\end{equation}

\noindent in which $p$ and $q$ are given in Eq.~\eqref{eq:the_parameters}.
The constraint \( 0.5 \leq \theta \leq 1 \) follows the settings used in prior work~\cite{tianhao2017}.

\subsection{Empirical Validation of Analytical ASR-MSE Frontiers} \label{sub:empirical_results}

To validate the analytical ASR-MSE framework introduced in Section~\ref{sub:asr_mse_framework}, we empirically verify that the closed-form ASR and MSE expressions produce Pareto frontiers consistent with the observed behavior on real data.
Concretely, we use the \textbf{Adult} dataset from the UCI machine learning repository~\cite{uci}, selecting the \texttt{Age} attribute with domain size $k=100$ (\ie, $\text{Age}\in[0,99]$) and $n=48842$ users, and report results averaged over $100$ independent runs.
This experiment serves as a consistency check of the analytical methodology.

\begin{figure*}[!htb]
    \centering
    \includegraphics[width=1\linewidth]{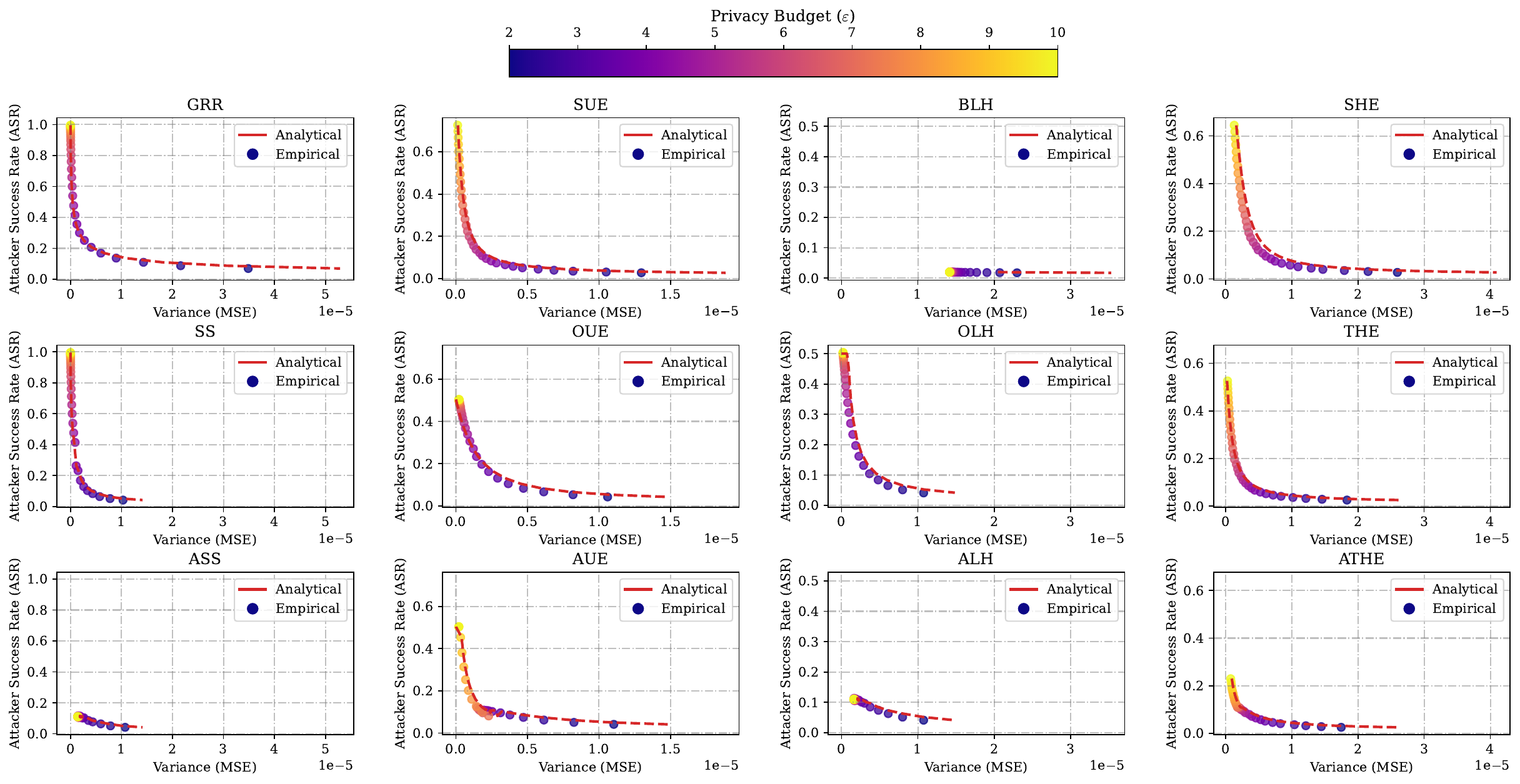}
    \caption{
    Comparison of empirical and analytical Pareto frontiers for ASR \vs{} Variance (MSE) across various LDP protocols (state-of-the-art and adaptive). 
    Each subplot considers a range of privacy budgets $\varepsilon \in (2, 10)$ and a fixed domain size $k=100$. 
    Empirical results ($\CIRCLE$ markers) are averaged over $100$ independent runs with the \texttt{Age} attribute of the Adult dataset~\cite{uci}, while analytical results (\textcolor{red}{red} dashed lines) are computed using closed-form equations.}
    \label{fig:fig_asr_mse_analytical_vs_empirical}
\end{figure*}

Figure~\ref{fig:fig_asr_mse_analytical_vs_empirical} presents the comparison between \textbf{empirical} ($\circ$ markers) and \textbf{analytical} (\textcolor{red}{red} dashed lines) Pareto frontiers for ASR versus MSE across various LDP protocols, including both state-of-the-art and our adaptive variants. 
Each subplot represents a specific protocol (\ie, GRR, SUE, BLH, SHE, SS, OUE, OLH, THE, ASS, AUE, ALH and ATHE) with the privacy budget varying as $\varepsilon \in (2, 10)$ with the domain size fixed at $k=100$. 
One can observe that across all protocols, the empirical results closely align with the analytical calculations, demonstrating the robustness of the closed-form ASR and MSE expressions.
This alignment highlights the reliability of our theoretical framework and previous analytical experiments across different privacy budgets and data distributions.  

\begin{leftbar}
\noindent \textbf{Analytical \vs{} empirical validation:} 
Our findings demonstrate a strong agreement between analytical and empirical ASR-MSE Pareto frontiers, supporting the use of closed-form expressions as a reliable basis for both systematic analysis and attack-aware LDP protocol design.
\end{leftbar}

\section{Experiments and Analysis} \label{sec:results}

The objective of our experiments is to thoroughly evaluate the performance of the proposed adaptive LDP protocols in terms of privacy, utility and resilience against adversarial threats. 
Specifically, in Section~\ref{sub:asr_results}, we conduct an \textbf{\emph{ASR analysis}} for each LDP protocol to quantify their vulnerability to data reconstruction attacks, thereby assessing the privacy guarantees offered by existing and newly proposed mechanisms.
Subsequently, in Section~\ref{sub:mse_results}, we perform an \textbf{\emph{MSE analysis}} to evaluate the utility guarantees provided by existing and our newly proposed mechanisms.
Next, in Section~\ref{sub:asr_mse_results}, we explore the \textbf{\emph{trade-off between ASR and MSE}} (\ie, Pareto frontier), highlighting how our adaptive protocols compare to traditional ones in balancing privacy and utility under different scenarios. 
Finally, in Section~\ref{sub:results_poisoning}, we extend our evaluation to a joint view of privacy leakage, estimation utility and integrity under data poisoning attacks.

To systematize these evaluations, we rely on \textbf{analytical closed-form expressions}, enabling a comprehensive exploration of the \emph{privacy-utility-attackability} trade-off across privacy budgets and domain sizes at low computational cost.
This choice is supported by prior work showing close agreement between analytical and empirical results~\cite{tianhao2017,Gursoy2022,Arcolezi_rs_fd}, and further validated by our empirical study in Section~\ref{sub:empirical_results}.

\subsection{Setup of Analytical Experiments} \label{sub:setup_experiments}

For all experiments, we have used the following settings:

\begin{itemize}
    \item \textbf{LDP protocols.} We experiment with the eight LDP protocols described in Section~\ref{sec:ldp_protocols} and our four adaptive LDP protocols described in Section~\ref{sub:our_protocols}.
    
    \item \textbf{Number of users.} For the analytical variance/MSE derived from Eq.~\eqref{eq:approx_var} (\eg, Eq.~\eqref{eq:var_grr} for GRR, Eq.~\eqref{eq:var_the} for THE, \ldots), we report variance per user, corresponding to the analytical derivation of \( \text{Var}[] / n \). 
    This allows us to examine the fundamental properties of each protocol independently of the dataset size.
    
    \item \textbf{Privacy parameter.} The LDP frequency estimation protocols were evaluated under two privacy regimes: 
        \begin{enumerate}[label=\alph*)]
            \item \textbf{High privacy regime} with $\varepsilon \in \{0.1, 0.2, \ldots, 1.9, 2.0\}$.
            \item \textbf{Medium to low privacy regime} with $\varepsilon \in \{2.0, 2.5, \ldots, 9.5, 10.0\}$.
        \end{enumerate}  

    \item \textbf{Domain size.} We vary the domain size in three ranges: 
        \begin{enumerate}[label=\alph*)]
            \item \textbf{Small domain:} $k \in \{25, 50, 75, 100\}$.
            \item \textbf{Medium domain:} $k \in \{250, 500, 750, 1000\}$.
            \item \textbf{Large domain:} $k \in \{2500, 5000, 7500, 10000\}$.
        \end{enumerate}

    \item \textbf{Weights for optimization.} 
    Unless otherwise mentioned, we fix the weights for the two-objective optimization in Eq.~\eqref{eq:objective_function} to \(w_{\text{ASR}}=w_{\mathrm{MSE}}=0.5\), aiming for a balanced trade-off between privacy and accuracy. 
    Experiments in Appendix~\ref{app:weights_results} focus on varying these weights.

    \item \textbf{Optimization method.} Parameters for our adaptive protocols (\eg, subset size \(\omega\) for ASS) were tuned via grid search~\cite{bergstra2012random}, followed by selection using a \textit{Weighted Scalarization} strategy (see Eq.~\eqref{eq:objective_function}).  
    While grid search is exhaustive and ensures broad coverage of the parameter space, it can be computationally expensive. 
    For large or high-dimensional search spaces, we recommend constrained or gradient-free optimization methods~\cite{brent2013algorithms} as a more efficient alternative to accelerate convergence without significantly sacrificing solution quality.
    Experiments in Appendix~\ref{app:parameter_results} analyze parameter selection.
\end{itemize}

\subsection{ASR Analysis for LDP Protocols} \label{sub:asr_results}

\begin{figure*}[t]
    \centering
    \includegraphics[width=0.975\linewidth]{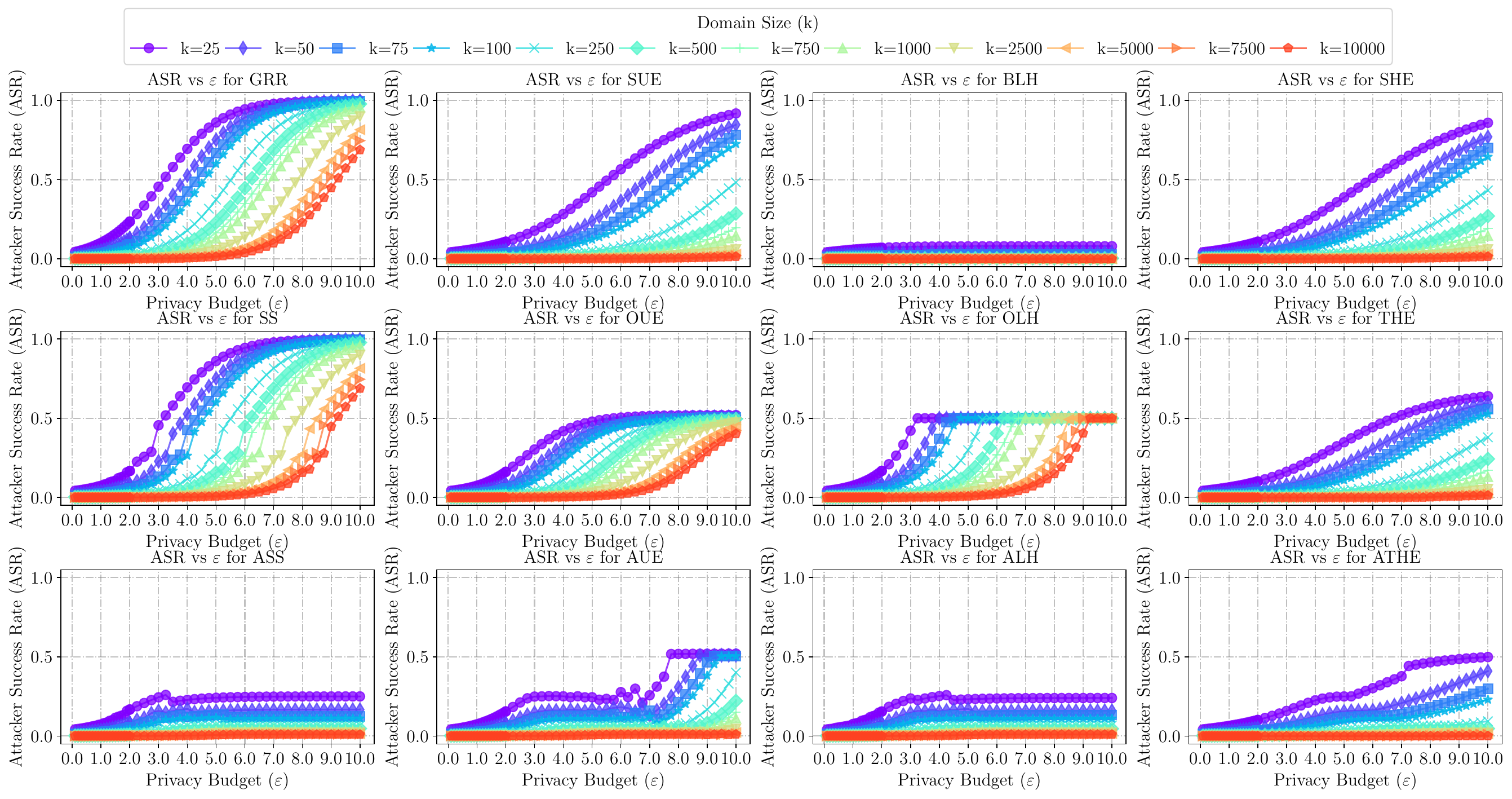}
    \caption{Attacker Success Rate (ASR) \vs{} privacy budget (\(\varepsilon\)) for different LDP frequency estimation protocols across varying domain sizes (\(k\)). 
    The plots compare state-of-the-art LDP protocols, including GRR, SUE, BLH, OUE, OLH, SS, SHE and THE, against our newly proposed adaptive protocols (ASS, AUE, ALH and ATHE). 
    Each curve represents a different domain size, with \(k\) ranging from $25$ to $10000$. }
    \label{fig:asr_analysis}
\end{figure*}

In Figure~\ref{fig:asr_analysis}, we evaluate the ASR for various LDP frequency estimation protocols as a function of the privacy budget \(\varepsilon\) from high to low privacy regimes across small to big domain sizes \(k\). 
The goal of this analysis is to assess the robustness of each protocol against adversarial inference attacks under varying privacy levels and domain sizes. 
Specifically, we compare state-of-the-art protocols (GRR, SUE, BLH, OUE, OLH, SS, SHE and THE) against our proposed adaptive protocols (ASS, AUE, ALH and ATHE) to understand how effective these methods are at balancing privacy and utility.

For all protocols, we observe that the ASR generally increases with increasing privacy budget \(\varepsilon\) in Figure~\ref{fig:asr_analysis}. 
This behavior is expected, as higher values of \(\varepsilon\) correspond to weaker privacy guarantees, allowing the adversary to infer user data more effectively. 
For smaller domain sizes (\ie, \(k \leq 100\)), the ASR rises sharply, suggesting that the adversary's ability to correctly guess the user's input improves significantly as \(\varepsilon\) increases.
In contrast, larger domain sizes (\ie, \(k \geq 1000\)) show a more gradual increase in ASR, indicating that a larger domain inherently offers greater privacy protection. 
Still, \textbf{\textit{our adaptive methods effectively counteract privacy threats regardless of $k$ and $\varepsilon$}}, underscoring the flexibility and robustness offered by our two-objective optimization framework.

Our adaptive protocols (\ie, ASS, AUE, ALH and ATHE) demonstrate significantly lower ASR compared to their traditional counterparts across all privacy budgets. 
For example, ASS effectively mitigates the ASR vulnerability of the state-of-the-art SS by capping the $\mathrm{ASR}$ below 0.25, in contrast to the traditional SS, where the $\mathrm{ASR}$ gets to $1$ (\ie, the adversary can fully infer the user's value). 
Similarly, AUE consistently achieves a lower or comparable ASR relative to the state-of-the-art OUE protocol, highlighting the benefits of its adaptive parameter optimization. 
For ALH, the adaptive mechanism achieves a balanced compromise between the low ASR of BLH and the moderate ASR of OLH by dynamically optimizing the hash domain size, resulting in a substantial reduction in ASR compared to traditional OLH. 
Finally, ATHE demonstrates more resilience than THE across all domain sizes, thereby showcasing the effectiveness of our framework in enhancing privacy protection.

\begin{leftbar} 
\noindent \textbf{ASR increases with $\varepsilon$, but our adaptive protocols resist:} 
As the privacy budget $\varepsilon$ increases, ASR generally rises for all protocols due to weaker privacy guarantees. 
However, our adaptive protocols (ASS, AUE, ALH and ATHE) exhibit significantly lower (\ie, by up to $\leq 5\times$) ASRs across a broad range of $\varepsilon$ values, underscoring their enhanced resilience against privacy attacks.
\end{leftbar}

\subsection{MSE Analysis for LDP Protocols} \label{sub:mse_results}

\begin{figure}[!htb]
    \centering
    \includegraphics[width=1\linewidth]{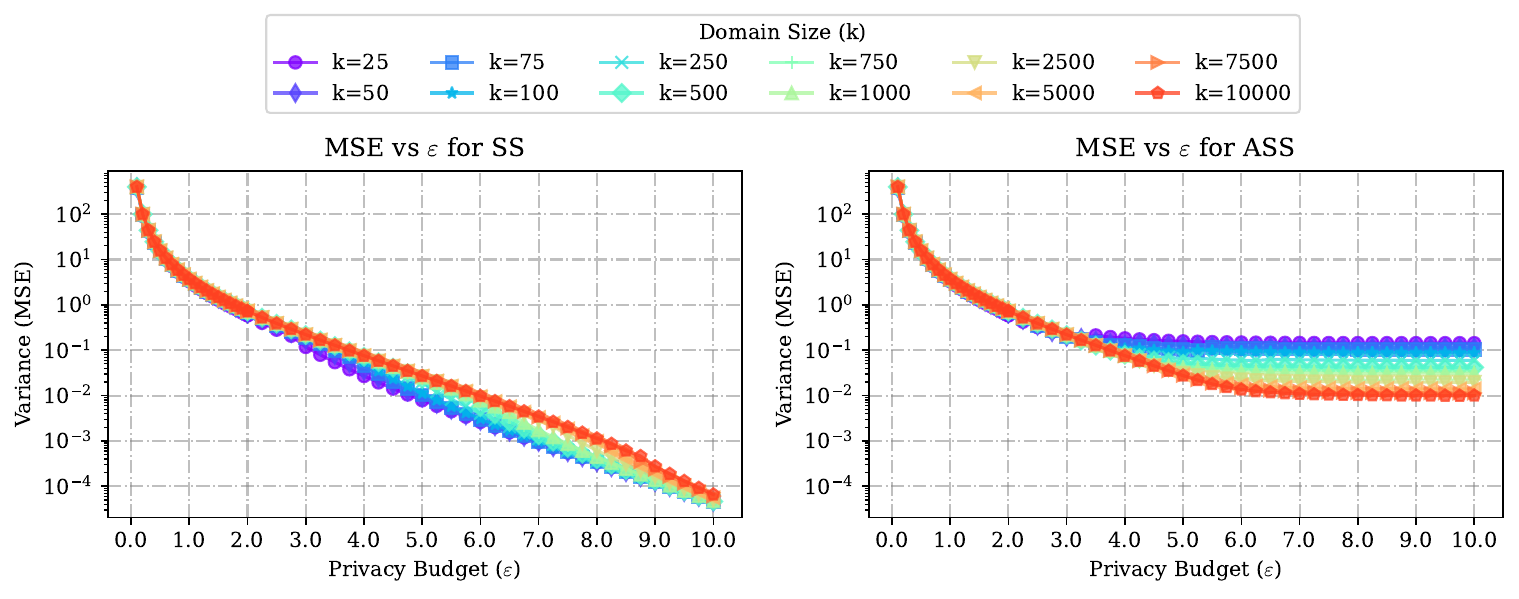}
    \caption{Variance (MSE) \vs{} privacy budget (\(\varepsilon\)) for the state-of-the-art SS protocol and our adaptive version ASS across various domain sizes \(k\). 
    Each curve represents a distinct domain size, illustrating how each protocol balances estimation accuracy with privacy as \(\varepsilon\) changes.}
    \label{fig:mse_analysis_SS_vs_ASS}
\end{figure}

\begin{figure*}[!htb]
    \centering
    \includegraphics[width=0.87\linewidth]{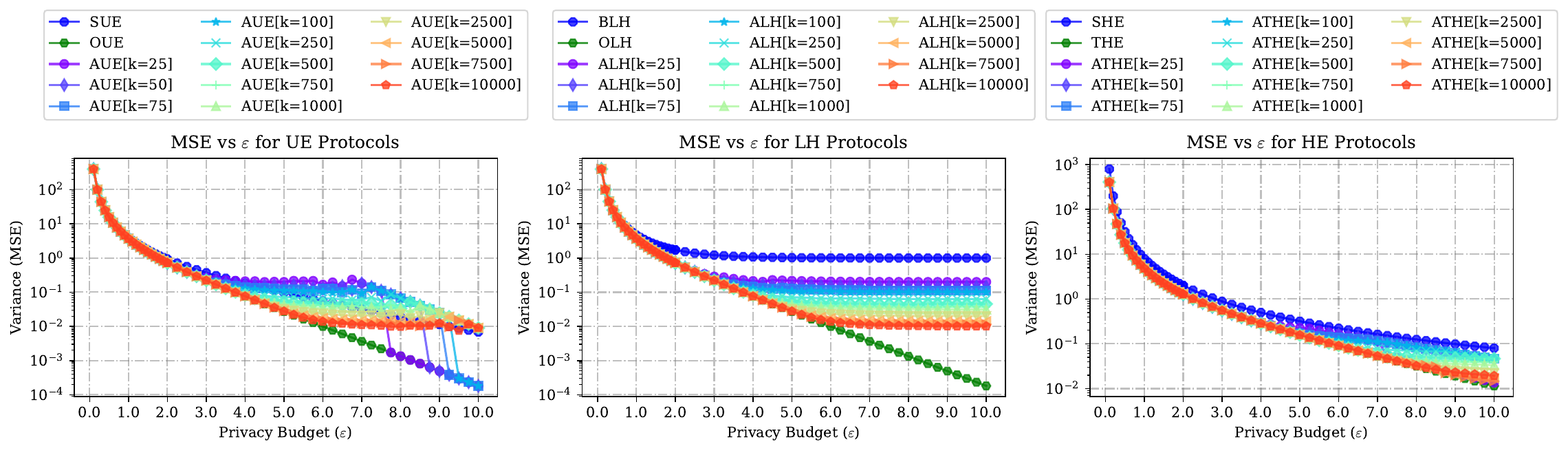}
    \caption{Variance (MSE) \vs{} privacy budget (\(\varepsilon\)) for the state-of-the-art LDP protocols (UE-, LH-, and HE-based) and our adaptive versions (AUE, ALH, and ATHE) across various domain sizes \(k\). 
    For our adaptive protocols, each curve represents a distinct domain size, illustrating how each protocol balances estimation accuracy with privacy as \(\varepsilon\) changes.}
    \label{fig:mse_analysis_other_UE_LH_HE}
\end{figure*}

In Figure~\ref{fig:mse_analysis_SS_vs_ASS}, we examine the MSE behavior of the SS protocol and our adaptive protocol ASS, across varying privacy budgets \(\varepsilon\). 
In Figure~\ref{fig:mse_analysis_other_UE_LH_HE}, we extend this analysis to compare MSE trends for UE-, LH- and HE-based LDP protocols, including our adaptive versions (AUE, ALH and ATHE). 
These analyses aim to determine whether the adaptive protocols' improved robustness to privacy attacks results in substantial estimation accuracy loss or if they maintain competitive MSE values.
Notice that the MSE curves of SUE, OUE, BLH, OLH, SHE and THE remain independent of the domain size $k$ due to their fixed parameterization, as established in~\cite{tianhao2017}. 

We observe in Figure~\ref{fig:mse_analysis_SS_vs_ASS} that ASS exhibits an increase in MSE by up to two orders of magnitude compared to SS, particularly in higher privacy regimes (\(\varepsilon \geq 4\)). 
This increase is expected, as SS's minimal MSE comes at the cost of extreme vulnerability, with ASR approaching 1 (\eg, see Figure~\ref{fig:asr_analysis}), meaning an adversary can fully infer the user's value. 
ASS, in contrast, balances this trade-off by introducing adaptive parameterization that mitigates privacy attacks while moderately increasing the MSE. 
In Figure~\ref{fig:mse_analysis_other_UE_LH_HE}, we observe similar trends for the other adaptive protocols (AUE, ALH, and ATHE). 
For UE protocols, our adaptive AUE version achieves slightly higher MSE compared to OUE across all domain sizes \(k\), with the gap becoming more pronounced as \(\varepsilon\) increases. 
This is expected, as AUE optimizes its parameters to enhance robustness to privacy attacks, leading to a slight trade-off in utility. 
Notably, AUE remains competitive with SUE in high privacy regimes (\(\varepsilon \leq 2\)), offering similar MSE levels while achieving significantly improved ASR performance as shown earlier. 
In low privacy regimes (\(\varepsilon > 2\)), AUE incurs a modest increase in MSE (1 order of magnitude) compared to OUE. 

Moreover, for LH- and HE-based protocols, we observe in Figure~\ref{fig:mse_analysis_other_UE_LH_HE} that our adaptive protocols (ALH and ATHE) demonstrate a variance (MSE) behavior that is consistently ``sandwiched'' between the two corresponding state-of-the-art protocols in their respective groups. 
Specifically, ALH achieves MSE values between BLH (which minimizes ASR at the cost of higher MSE) and OLH (which minimizes MSE but is more vulnerable to privacy attacks). 
Similarly, ATHE's variance lies between SHE and THE, showing a trade-off where ATHE retains competitive MSE while prioritizing adversarial resilience. 
This positioning highlights how adaptivity allows our protocols to achieve a better privacy-utility trade-off.

\begin{leftbar}
\noindent \textbf{Adaptive protocols maintain competitive MSE:} 
Our adaptive protocols (ASS, AUE, ALH and ATHE) achieve higher MSE compared to their non-adaptive counterparts.
However, these increases remain within acceptable bounds ($\leq2$ orders of magnitude), highlighting that the improved adversarial resilience (ASR $\leq5 \times$) comes at a reasonable cost to the utility.
This suggests that enhanced privacy protection against adversaries can be achieved without prohibitive increases in variance.
\end{leftbar}

\subsection{Pareto Frontier for ASR and MSE}
\label{sub:asr_mse_results}

\begin{figure*}[htb]
    \centering
    \includegraphics[width=1\linewidth]{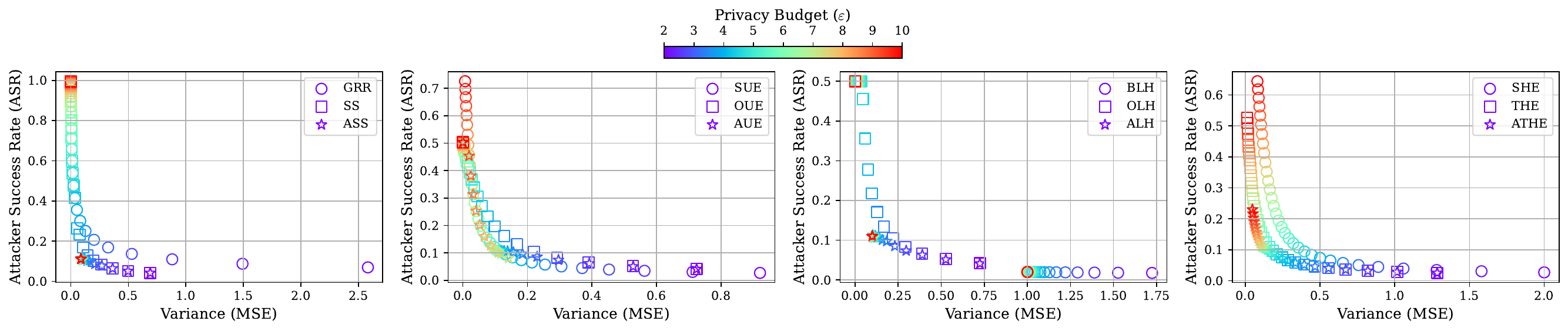}
    \caption{
    Attacker Success Rate (ASR) \vs{} Variance (MSE) for LDP frequency estimation protocols under data reconstruction attacks, with fixed domain size $k=100$.
    Each point corresponds to a distinct privacy budget $\varepsilon$ (medium to low privacy regimes), with colors indicating the privacy level.
    Protocols are organized by encoding family, comparing state-of-the-art mechanisms (GRR, SUE, BLH, SHE, SS, OUE, OLH, THE) against our adaptive counterparts (ASS, AUE, ALH, ATHE).}
    \label{fig:asr_mse_tradeoff_k100}
\end{figure*}

Beyond analyzing ASR and MSE independently as functions of the privacy budget $\varepsilon$ (see Figures~\ref{fig:asr_analysis},~\ref{fig:mse_analysis_other_UE_LH_HE} and~\ref{fig:mse_analysis_SS_vs_ASS}), it is crucial to examine how privacy leakage and estimation accuracy interact jointly.
To this end, we study the Pareto frontier between attackability (ASR) and estimation error (MSE).
Figure~\ref{fig:asr_mse_tradeoff_k100} reports ASR \vs{} MSE for a fixed domain size $k=100$, with the privacy budget $\varepsilon$ varying in the medium-to-low privacy regime.
Protocols are grouped by their encoding family (GRR/SS, UE, LH and HE), allowing for direct comparison between state-of-the-art mechanisms and their adaptive counterparts under identical domain and privacy conditions.
Each point corresponds to a specific value of $\varepsilon$, with colors indicating the privacy level.
Additional \emph{per-protocol} ASR-MSE trade-offs for other domain sizes and privacy regimes are provided in Appendix~\ref{app:add_asr_mse_tradeoff}.

From Figure~\ref{fig:asr_mse_tradeoff_k100}, comparisons within each encoding family reveal systematic structural differences in how protocols navigate the ASR-MSE trade-off.
GRR and SS exhibit steep trajectories: as $\varepsilon$ increases, they rapidly reach low-MSE regimes at the cost of sharply increasing ASR, indicating strong utility but severe privacy leakage.
In contrast, SUE-, SHE- and THE-based mechanisms follow smoother ASR-MSE paths, yielding more gradual trade-offs and avoiding abrupt transitions into highly vulnerable regions.
OUE and OLH achieve low variance primarily at higher $\varepsilon$ values, where ASR plateaus around $0.5$, reflecting their utility-oriented parameterization.
By contrast, BLH consistently maintains low ASR across the entire range of $\varepsilon$, but does so at the expense of substantially higher MSE, illustrating a conservative privacy-first design.
Notably, because OUE is designed to minimize MSE, SUE achieves a more favorable ASR-MSE trade-off across a broad range of privacy budgets.
This illustrates that minimizing estimation error alone does not necessarily yield the best privacy-utility balance under reconstruction attacks.

Our adaptive protocols fundamentally reshape these within-family trade-offs.
ASS mitigates the sharp ASR escalation observed in SS at low-MSE operating points.
AUE improves upon OUE by maintaining a lower ASR without incurring disproportionate variance increases, particularly in moderate privacy regimes, and in several regimes, it overlaps with SUE in the ASR-MSE space.
Likewise, ALH and ATHE consistently dominate their non-adaptive counterparts by jointly reducing ASR without incurring excessive variance increases.
Rather than interpolating between existing designs, these adaptive mechanisms shift the ASR-MSE trade-off frontier itself, enabling operating points that are unattainable through varying $\varepsilon$ alone under fixed protocol parameterizations.

\begin{leftbar}
\noindent \textbf{Key insight:} 
Each protocol exhibits a distinct ASR-MSE profile. By jointly optimizing for both ASR and MSE, our adaptive protocols consistently push these curves toward more favorable regimes in the ASR-MSE Pareto frontier.
\end{leftbar}

\subsection{Complementary Adversarial Perspective: Poisoning Attacks}
\label{sub:results_poisoning}

While ASR captures vulnerability to data reconstruction attacks, LDP protocols may also be susceptible to \emph{data poisoning attacks}~\cite{cao2021data}, in which an adversary injects malicious reports to bias the aggregated frequency estimates.
To capture this integrity-oriented threat (see Eq.~\eqref{eq:multi_objective_framework}), we adopt the \emph{maximal gain attack} (MGA) proposed by Cao \emph{et al.}~\cite{cao2021data}, which is shown to be the strongest poisoning strategy in terms of maximizing the expected increase in the estimated total frequency of a target item set $T$.
For UE and LH protocols, the MGA admits a unified closed-form expression, referred to as the attacker’s \emph{Expected Poisoning Gain (EPG)}:
\begin{equation}
\label{eq:epg_pq}
\mathrm{EPG}
=
\beta\!\left(
\frac{r(1-q)}{p-q}
-
f_T
\right),
\end{equation}
in which $\beta$ denotes the fraction of fake users, $r = |T|$ is the number of targeted items and $f_T$ is the total true frequency mass of these items among genuine users.
Following the experimental setup of~\cite{cao2021data}, we fix the poisoning parameters to $\beta = 0.05$, $r = 1$, and $f_T = 0.01$, ensuring comparability and avoiding additional attacker-side tuning degrees of freedom.

\begin{figure}[!htb]
    \centering
    \begin{subfigure}[t]{0.88\linewidth}
        \centering
        \includegraphics[width=\linewidth]{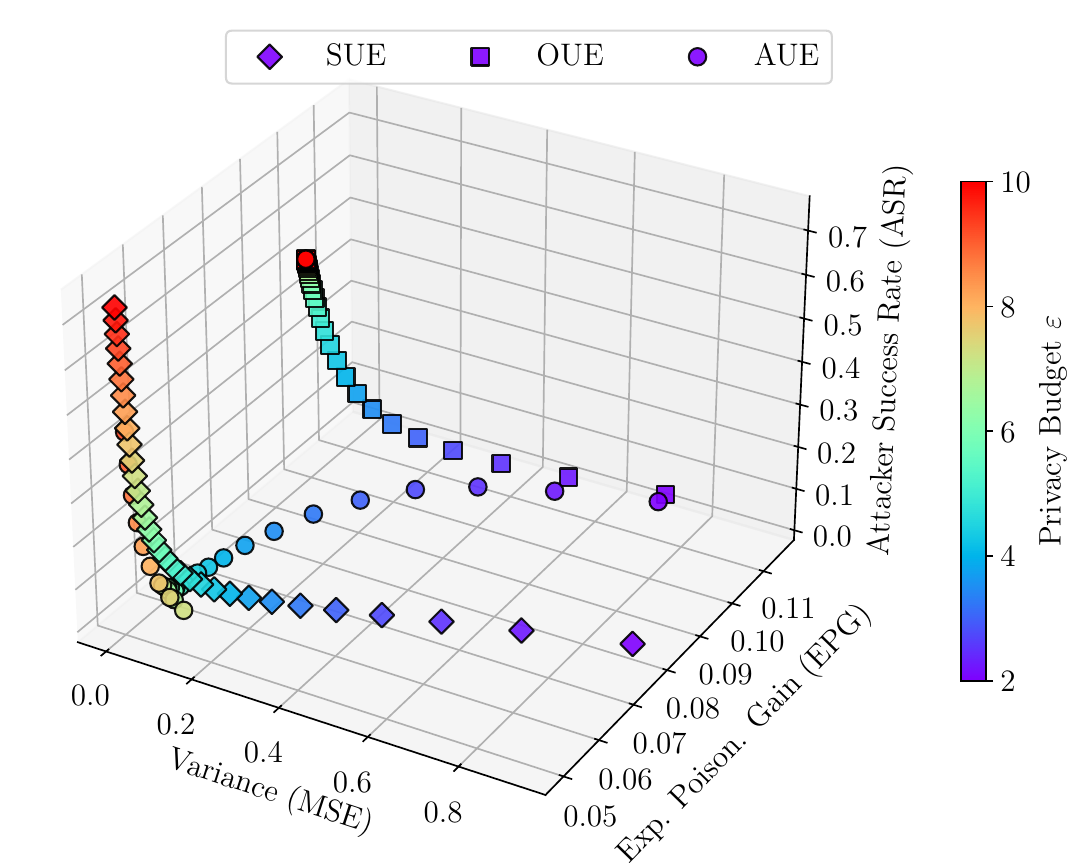}
        \caption{UE-based protocols: SUE, OUE, and AUE.}
        \label{fig:asr_mse_epg_ue}
    \end{subfigure}
    \hfill\\
    \begin{subfigure}[t]{0.88\linewidth}
        \centering
        \includegraphics[width=\linewidth]{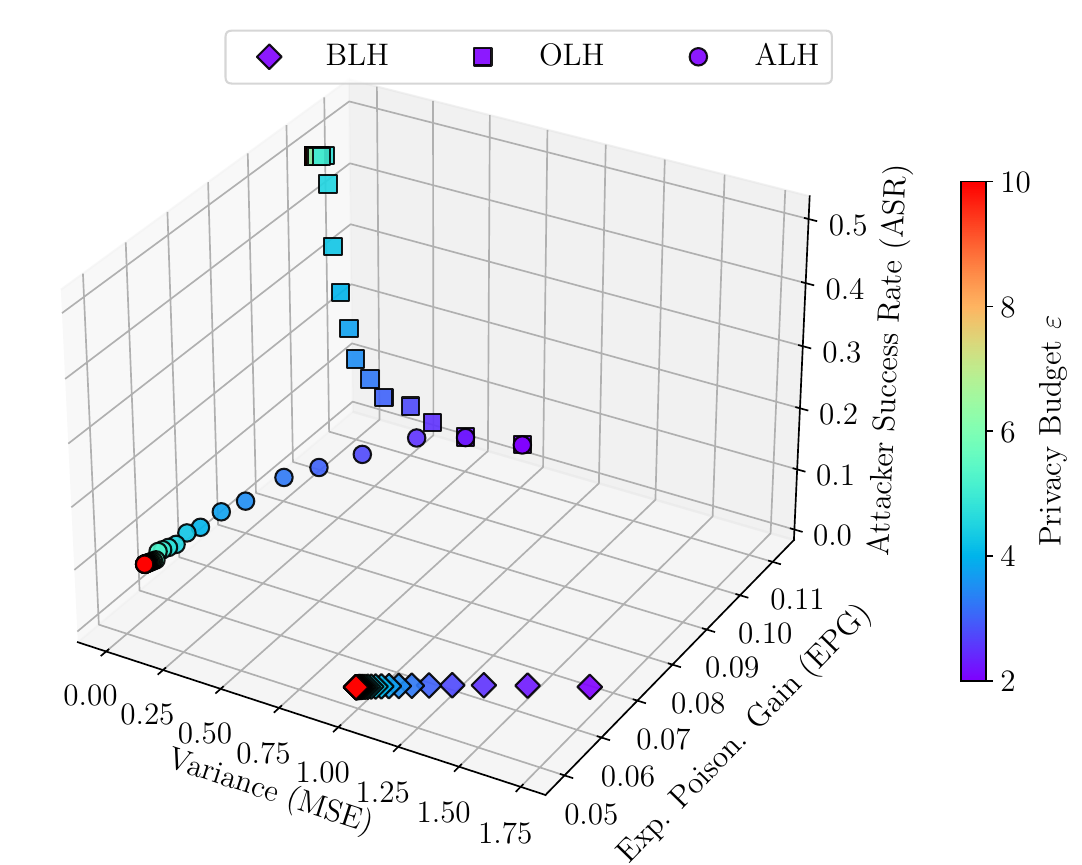}
        \caption{LH-based protocols: BLH, OLH, and ALH.}
        \label{fig:asr_mse_epg_lh}
    \end{subfigure}

    \caption{
    Joint ASR-MSE-EPG trade-off for UE- and LH-based LDP frequency estimation protocols at domain size $k=100$.
    Each point corresponds to a distinct privacy budget $\varepsilon \in (2,10)$.}
    \label{fig:asr_mse_epg_3d}
\end{figure}

\textbf{Joint View.} 
Taken together, ASR, EPG and MSE define a three-dimensional trade-off between privacy leakage, estimation utility and integrity robustness.
Figure~\ref{fig:asr_mse_epg_3d} provides a geometric illustration of this joint space for state-of-the-art UE- and LH-based protocols and our adaptive counterparts, where each point corresponds to a distinct $\varepsilon \in (2,10)$ and a fixed $k=100$.
A key implication is that decreasing $\varepsilon$ simultaneously increases the estimation error (MSE) and poisoning vulnerability (EPG), while reducing the reconstruction risk (ASR).
Rather than prescribing a single operating point, this three-objective view supports multi-objective optimization or projections onto lower-dimensional trade-offs~\cite{Donoso2007} (\eg, ASR \vs{} MSE).
In our experiments, adaptive protocols are tuned exclusively with respect to ASR and MSE (Section~\ref{sub:our_protocols}), while EPG is evaluated \emph{post hoc} to assess the resulting integrity implications.
More generally, alternative adaptive designs could directly incorporate poisoning robustness into the optimization objective, \eg, through weighted combinations of ASR, MSE and EPG as Eq.~\eqref{eq:objective_function}, which we leave as future work.

Within this three-dimensional space, our adaptive protocols (AUE and ALH) reshape the attainable trade-offs at fixed $\varepsilon$.
For UE-based protocols, AUE systematically bridges the gap between the utility-oriented OUE and the more conservative SUE, achieving comparable reconstruction rates to SUE at substantially lower estimation error.
For LH-based protocols, ALH reduces ASR relative to OLH and, as a consequence of its parameter adaptation, also yields lower EPG, while achieving robustness levels comparable to BLH with substantially improved utility.
Overall, these results demonstrate that attack-aware parameter adaptation reveals operating points that cannot be achieved by tuning the privacy budget alone.

\section{Conclusion and Perspectives} \label{sec:conclusion}

This work presented a comprehensive experimental and analytical benchmark of LDP frequency estimation protocols, emphasizing the joint evaluation of privacy leakage, utility, and integrity under adversarial settings. 
By positioning existing protocols within a unified \emph{privacy-utility-attackability} trade-off space, we provided a reproducible methodology to assess and compare their resilience against both data reconstruction and poisoning attacks.
Our study revealed clear and protocol-dependent Pareto frontiers between ASR under \textit{data reconstruction attacks}~\cite{Gursoy2022,Arcolezi2023,arcolezi2024revealing} and MSE, as well as their interaction with integrity-oriented attackability, exposing vulnerabilities not captured by $\varepsilon$ alone.

Beyond analysis, we showed that systematic benchmarking can guide the refinement of existing mechanisms. 
By leveraging multi-objective optimization, we derived adaptive parameterizations (ASS, AUE, ALH, ATHE) that achieve more favorable privacy-utility-attackability trade-offs than their original counterparts (SS, OUE, OLH, THE) while strictly preserving $\varepsilon$-LDP guarantees. 
These adaptive variants demonstrate that informed tuning can substantially enhance robustness against adversarial threats without prohibitive utility loss.

Overall, our work establishes a principled approach to evaluating LDP mechanisms through standardized, attack-aware benchmarks that jointly capture privacy, utility and integrity considerations. 
We envision this benchmark as a foundation for a broader suite of experimental tools for LDP, enabling future studies to extend our evaluation to additional adversarial models (\eg, attribute inference~\cite{Gadotti2022,Arcolezi2023evolving,GURSOY2024}, re-identification~\cite{Murakami2021,Arcolezi2023}) and complementary system-level metrics such as communication and computation cost.

\clearpage

\section*{Acknowledgments}
\noindent 
The authors thank the anonymous reviewers for their insightful suggestions and Haoying Zhang for her helpful comments on an earlier draft.
Héber H. Arcolezi has been partially supported by the French National Research Agency (ANR), under contracts: ``ANR-24-CE23-6239'' and ``ANR-23-IACL-0006''.
Sébastien Gambs is supported by the Canada Research Chair program as well as a Discovery Grant from NSERC. 

\section*{AI-Generated Content Acknowledgement}
\noindent The authors acknowledge the use of ChatGPT (OpenAI, GPT-5 model) to assist with language-related improvements, including grammar correction, spelling, formatting consistency and refinement of phrasing for clarity and readability. 
No part of the paper’s scientific content, analysis, experimental design, results or conclusions was generated or influenced by AI tools. 
All conceptual, technical and experimental contributions originate solely from the authors.

\bibliographystyle{ieeetr}
\bibliography{B_references}

\begin{thebibliography}{10}

\bibitem{Dwork2006}
C.~Dwork, F.~McSherry, K.~Nissim, and A.~Smith, ``Calibrating noise to sensitivity in private data analysis,'' in {\em Theory of Cryptography}, pp.~265--284, Springer Berlin Heidelberg, 2006.

\bibitem{first_ldp}
S.~P. Kasiviswanathan, H.~K. Lee, K.~Nissim, S.~Raskhodnikova, and A.~Smith, ``What can we learn privately?,'' {\em SIAM Journal on Computing}, vol.~40, no.~3, pp.~793--826, 2011.

\bibitem{rappor}
U.~Erlingsson, V.~Pihur, and A.~Korolova, ``{RAPPOR}: Randomized aggregatable privacy-preserving ordinal response,'' in {\em Proceedings of the 2014 ACM SIGSAC Conference on Computer and Communications Security}, (New York, NY, USA), pp.~1054--1067, ACM, 2014.

\bibitem{apple}
\textrm{Apple Differential Privacy Team}, ``Learning with privacy at scale,'' 2017.
\newblock \url{https://docs-assets.developer.apple.com/ml-research/papers/learning-with-privacy-at-scale.pdf}.

\bibitem{microsoft}
B.~Ding, J.~Kulkarni, and S.~Yekhanin, ``Collecting telemetry data privately,'' in {\em Advances in Neural Information Processing Systems 30} (I.~Guyon, U.~V. Luxburg, S.~Bengio, H.~Wallach, R.~Fergus, S.~Vishwanathan, and R.~Garnett, eds.), pp.~3571--3580, Curran Associates, Inc., 2017.

\bibitem{Bassily2015}
R.~Bassily and A.~Smith, ``Local, private, efficient protocols for succinct histograms,'' in {\em Proceedings of the Forty-Seventh Annual ACM Symposium on Theory of Computing}, STOC '15, (New York, NY, USA), p.~127–135, Association for Computing Machinery, 2015.

\bibitem{Li2024}
X.~Li, W.~Liu, J.~Lou, Y.~Hong, L.~Zhang, Z.~Qin, and K.~Ren, ``Local differentially private heavy hitter detection in data streams with bounded memory,'' {\em Proc. ACM Manag. Data}, vol.~2, Mar. 2024.

\bibitem{Zhang2025}
Y.~Zhang, Q.~Ye, and H.~Hu, ``Federated heavy hitter analytics with local differential privacy,'' {\em Proc. ACM Manag. Data}, vol.~3, Feb. 2025.

\bibitem{Arcolezi2022}
H.~H. Arcolezi, J.-F. Couchot, B.~A. Bouna, and X.~Xiao, ``Improving the utility of locally differentially private protocols for longitudinal and multidimensional frequency estimates,'' {\em Digital Communications and Networks}, vol.~10, no.~2, pp.~369--379, 2024.

\bibitem{Arcolezi2023evolving}
H.~H. Arcolezi, C.~A. Pinzón, C.~Palamidessi, and S.~Gambs, ``Frequency estimation of evolving data under local differential privacy,'' in {\em Proceedings of the 26th International Conference on Extending Database Technology, {EDBT} 2023, Ioannina, Greece, March 28 - March 31, 2023}, pp.~512--525, OpenProceedings.org, 2023.

\bibitem{Gu2019}
X.~Gu, M.~Li, Y.~Cao, and L.~Xiong, ``Supporting both range queries and frequency estimation with local differential privacy,'' in {\em 2019 IEEE Conference on Communications and Network Security (CNS)}, pp.~124--132, 2019.

\bibitem{yang2020answering}
J.~Yang, T.~Wang, N.~Li, X.~Cheng, and S.~Su, ``Answering multi-dimensional range queries under local differential privacy,'' {\em arXiv preprint arXiv:2009.06538}, 2020.

\bibitem{Arcolezi_rs_fd}
H.~H. Arcolezi, J.-F. Couchot, B.~Al~Bouna, and X.~Xiao, ``Random sampling plus fake data: Multidimensional frequency estimates with local differential privacy,'' in {\em Proceedings of the 30th {ACM} International Conference on Information \& Knowledge Management}, pp.~47--57, {ACM}, oct 2021.

\bibitem{Filho2023}
J.~S. Costa~Filho and J.~C. Machado, ``Felip: A local differentially private approach to frequency estimation on multidimensional datasets,'' in {\em Proceedings of the 26th International Conference on Extending Database Technology, {EDBT} 2023, Ioannina, Greece, March 28 - March 31, 2023}, pp.~671--683, OpenProceedings.org, 2023.

\bibitem{Li2022}
J.~Li, W.~Gan, Y.~Gui, Y.~Wu, and P.~S. Yu, ``Frequent itemset mining with local differential privacy,'' in {\em Proceedings of the 31st ACM International Conference on Information \& Knowledge Management}, CIKM '22, (New York, NY, USA), p.~1146–1155, Association for Computing Machinery, 2022.

\bibitem{Wu2023}
H.~Wu, R.~Ran, S.~Peng, M.~Yang, and T.~Guo, ``Mining frequent items from high-dimensional set-valued data under local differential privacy protection,'' {\em Expert Systems with Applications}, vol.~234, p.~121105, Dec. 2023.

\bibitem{alptekin2023building}
E.~Alptekin and M.~E. Gursoy, ``Building quadtrees for spatial data under local differential privacy,'' in {\em IFIP Annual Conference on Data and Applications Security and Privacy}, pp.~22--39, Springer, 2023.

\bibitem{Tire2024}
E.~Tire and M.~E. Gursoy, ``Answering spatial density queries under local differential privacy,'' {\em IEEE Internet of Things Journal}, vol.~11, no.~10, pp.~17419--17436, 2024.

\bibitem{Raab2025}
R.~Raab, P.~Berrang, P.~Gerhart, and D.~Schr\"{o}der, ``Sok: Descriptive statistics under local differential privacy,'' {\em Proceedings on Privacy Enhancing Technologies}, vol.~2025, p.~118–149, Jan. 2025.

\bibitem{kairouz2016discrete}
P.~Kairouz, K.~Bonawitz, and D.~Ramage, ``Discrete distribution estimation under local privacy,'' in {\em International Conference on Machine Learning}, pp.~2436--2444, PMLR, 2016.

\bibitem{wang2016mutual}
S.~Wang, L.~Huang, P.~Wang, Y.~Nie, H.~Xu, W.~Yang, X.-Y. Li, and C.~Qiao, ``Mutual information optimally local private discrete distribution estimation,'' {\em arXiv preprint arXiv:1607.08025}, 2016.

\bibitem{Min2018}
M.~Ye and A.~Barg, ``Optimal schemes for discrete distribution estimation under locally differential privacy,'' {\em IEEE Transactions on Information Theory}, vol.~64, no.~8, pp.~5662--5676, 2018.

\bibitem{tianhao2017}
T.~Wang, J.~Blocki, N.~Li, and S.~Jha, ``Locally differentially private protocols for frequency estimation,'' in {\em 26th {USENIX} Security Symposium ({USENIX} Security 17)}, (Vancouver, BC), pp.~729--745, {USENIX} Association, Aug. 2017.

\bibitem{Hadamard}
J.~Acharya, Z.~Sun, and H.~Zhang, ``Hadamard response: Estimating distributions privately, efficiently, and with little communication,'' in {\em Proceedings of the Twenty-Second International Conference on Artificial Intelligence and Statistics} (K.~Chaudhuri and M.~Sugiyama, eds.), vol.~89 of {\em Proceedings of Machine Learning Research}, pp.~1120--1129, PMLR, 16--18 Apr 2019.

\bibitem{Feldman2022}
V.~Feldman, J.~Nelson, H.~Nguyen, and K.~Talwar, ``Private frequency estimation via projective geometry,'' in {\em Proceedings of the 39th International Conference on Machine Learning} (K.~Chaudhuri, S.~Jegelka, L.~Song, C.~Szepesvari, G.~Niu, and S.~Sabato, eds.), vol.~162 of {\em Proceedings of Machine Learning Research}, pp.~6418--6433, PMLR, 17--23 Jul 2022.

\bibitem{arcolezi2025private}
H.~H. Arcolezi, ``Private frequency estimation via residue number systems,'' {\em arXiv preprint arXiv:2511.11569}, 2025.

\bibitem{Murakami2021}
T.~Murakami and K.~Takahashi, ``Toward evaluating re-identification risks in the local privacy model,'' {\em Transactions on Data Privacy}, vol.~14, no.~3, pp.~79--116, 2021.

\bibitem{Arcolezi2023}
H.~H. Arcolezi, S.~Gambs, J.-F. Couchot, and C.~Palamidessi, ``On the risks of collecting multidimensional data under local differential privacy,'' {\em Proc. VLDB Endow.}, vol.~16, p.~1126–1139, jan 2023.

\bibitem{Gadotti2022}
A.~Gadotti, F.~Houssiau, M.~S. M.~S. Annamalai, and Y.-A. de~Montjoye, ``Pool inference attacks on local differential privacy: Quantifying the privacy guarantees of apple{\textquoteright}s count mean sketch in practice,'' in {\em 31st USENIX Security Symposium (USENIX Security 22)}, (Boston, MA), pp.~501--518, USENIX Association, Aug. 2022.

\bibitem{Gursoy2022}
M.~Emre~Gursoy, L.~Liu, K.-H. Chow, S.~Truex, and W.~Wei, ``An adversarial approach to protocol analysis and selection in local differential privacy,'' {\em IEEE Transactions on Information Forensics and Security}, vol.~17, pp.~1785--1799, 2022.

\bibitem{arcolezi2024revealing}
H.~H. Arcolezi and S.~Gambs, ``Revealing the true cost of locally differentially private protocols: An auditing perspective,'' {\em Proceedings on Privacy Enhancing Technologies}, vol.~2024, no.~4, pp.~123--141, 2024.

\bibitem{wang2016using}
Y.~Wang, X.~Wu, and D.~Hu, ``Using randomized response for differential privacy preserving data collection,'' in {\em EDBT/ICDT Workshops}, vol.~1558, pp.~0090--6778, 2016.

\bibitem{Cormode2021}
G.~Cormode, S.~Maddock, and C.~Maple, ``Frequency estimation under local differential privacy,'' {\em Proceedings of the {VLDB} Endowment}, vol.~14, pp.~2046--2058, July 2021.

\bibitem{cao2021data}
X.~Cao, J.~Jia, and N.~Z. Gong, ``Data poisoning attacks to local differential privacy protocols,'' in {\em 30th {USENIX} Security Symposium ({USENIX} Security 21)}, pp.~947--964, USENIX Association, Aug. 2021.

\bibitem{GURSOY2024}
M.~E. G\"{U}RSOY, ``Longitudinal attacks against iterative data collection with local differential privacy,'' {\em Turkish Journal of Electrical Engineering and Computer Sciences}, vol.~32, p.~198–218, Feb. 2024.

\bibitem{Balle2022}
B.~Balle, G.~Cherubin, and J.~Hayes, ``Reconstructing training data with informed adversaries,'' in {\em 2022 IEEE Symposium on Security and Privacy (SP)}, pp.~1138--1156, 2022.

\bibitem{Guerra2024}
P.~Guerra-Balboa, A.~Sauer, and T.~Strufe, ``Analysis and measurement of attack resilience of differential privacy,'' in {\em Proceedings of the 23rd Workshop on Privacy in the Electronic Society}, WPES '24, (New York, NY, USA), p.~155–171, Association for Computing Machinery, 2024.

\bibitem{Samarati1998}
P.~Samarati and L.~Sweeney, ``Generalizing data to provide anonymity when disclosing information (abstract),'' in {\em Proceedings of the Seventeenth ACM SIGACT-SIGMOD-SIGART Symposium on Principles of Database Systems}, PODS '98, p.~188, Association for Computing Machinery, 1998.

\bibitem{sun2024private}
Z.~Sun, P.~Kairouz, H.~Sun, A.~Gascon, and A.~T. Suresh, ``Private federated discovery of out-of-vocabulary words for gboard,'' {\em arXiv preprint arXiv:2404.11607}, 2024.

\bibitem{Warner1965}
S.~L. Warner, ``Randomized response: A survey technique for eliminating evasive answer bias,'' {\em Journal of the American Statistical Association}, vol.~60, pp.~63--69, Mar. 1965.

\bibitem{Cheu2021}
A.~Cheu, A.~Smith, and J.~Ullman, ``Manipulation attacks in local differential privacy,'' in {\em 2021 {IEEE} Symposium on Security and Privacy ({SP})}, {IEEE}, May 2021.

\bibitem{Barnes2020}
L.~P. Barnes, W.-N. Chen, and A.~Özgür, ``Fisher information under local differential privacy,'' {\em IEEE Journal on Selected Areas in Information Theory}, vol.~1, no.~3, pp.~645--659, 2020.

\bibitem{rao1992information}
C.~R. Rao, ``Information and the accuracy attainable in the estimation of statistical parameters,'' in {\em Breakthroughs in Statistics: Foundations and basic theory}, pp.~235--247, Springer, 1992.

\bibitem{CRAMÉR1999}
H.~Cramér, {\em Mathematical Methods of Statistics (PMS-9)}.
\newblock Princeton University Press, 1999.

\bibitem{Cover2006}
T.~M. Cover and J.~A. Thomas, {\em Elements of Information Theory (Wiley Series in Telecommunications and Signal Processing)}.
\newblock USA: Wiley-Interscience, 2006.

\bibitem{bergstra2012random}
J.~Bergstra and Y.~Bengio, ``Random search for hyper-parameter optimization.,'' {\em Journal of machine learning research}, vol.~13, no.~2, 2012.

\bibitem{Donoso2007}
Y.~Donoso and R.~Fabregat, {\em Multi-Objective Optimization in Computer Networks Using Metaheuristics}.
\newblock Auerbach Publications, 1~ed., 2007.

\bibitem{uci}
D.~Dua and C.~Graff, ``{UCI} machine learning repository,'' 2017.
\newblock Available online: \url{http://archive.ics.uci.edu/ml}.

\bibitem{brent2013algorithms}
R.~P. Brent, {\em Algorithms for minimization without derivatives}.
\newblock Courier Corporation, 2013.

\end{thebibliography}

\clearpage
\newpage
\onecolumn
\appendix

\subsection{Expected Data Reconstruction Attack Analyses}  \label{app:expected_asr}

\subsubsection{UE Protocols} \label{app:asr_ue}
Following the attack strategy for UE in Section~\ref{sub:ue_protocols}, we consider two events:

\begin{itemize}
    \item \textbf{Event 0:} The bit corresponding to the user's value \( x \) is flipped from 1 to 0, and all other bits remain 0.
    \begin{itemize}
        \item The attacker's guess is uniformly distributed over all \( k \) possible values.
       
        \item Success rate: \(\frac{1}{k}\).
       
        \item Probability: \(\Pr(\text{Event 0}) = (1 - p)(1 - q)^{k-1}\).
    \end{itemize}

    \item \textbf{Event 1:} The bit corresponding to the user's value \( x \) remains 1, and \( m-1 \) of the remaining \( k-1 \) bits are flipped from 0 to 1.
    \begin{itemize}
        \item The attacker's guess is uniformly distributed over the bits set to 1.
        
        \item If \( m \) bits are set to 1, the success rate is \(\frac{1}{m}\).
        
        \item Probability: \(\Pr(\text{Event 1 with } m \text{ bits set to 1}) = \binom{k-1}{m-1} p (q)^{m-1} (1 - q)^{k-m}\).
    \end{itemize}
\end{itemize}

Thus, combining these two events, we can derive the expected ASR as:
\[
\begin{aligned}
    \mathbb{E}[\text{ASR}]_{\mathrm{UE}} = &\Pr(\text{Event 0}) \cdot \frac{1}{k}\\ 
    &+ \sum_{m=1}^{k} \Pr(\text{Event 1 with } m \text{ bits set to 1}) \cdot \frac{1}{m} \mathrm{.}
\end{aligned}
\]

More formally, the probability calculations of each event are:

\begin{enumerate} 
    \item \textbf{Probability of Event 0:} 
    \[
        \Pr(\text{Event 0}) = (1 - p)(1 - q)^{k-1} \mathrm{.}
    \]

    \item \textbf{Probability of Event 1 with \( m \) bits set to 1:} 
    \[
    \begin{aligned}
        \Pr(\text{Event 1 with } m \text{ bits set to 1}) &= \\
         \binom{k-1}{m-1} p (q)^{m-1} (1 - q)^{k-m} &\mathrm{.}
    \end{aligned}
    \]
\end{enumerate}

Combining these probabilities, the expected ASR for UE is:
\begin{equation}\label{eq:exp_asr_ue_sum}
\begin{aligned}
    \mathbb{E}[\text{ASR}]_{\mathrm{UE}} = & \, (1 - p) \cdot (1 - q)^{k-1} \cdot \frac{1}{k} \\
    & + \sum_{m=1}^{k} p \cdot \frac{1}{m} \cdot \binom{k-1}{m-1} q^{m-1} (1-q)^{(k-1)-(m-1)} \mathrm{.}
\end{aligned}
\end{equation}

Since \(M=1+B\) with \(B\sim\mathrm{Bin}(k-1,q)\),

\[
\sum_{m=1}^{k}\frac{1}{m}\binom{k-1}{m-1}q^{\,m-1}(1-q)^{k-m}
=\mathbb{E}\!\left[\frac{1}{1+B}\right]
=\frac{1-(1-q)^k}{kq},
\]

Therefore, Equation~\eqref{eq:exp_asr_ue_sum} simplifies to:

\begin{equation}
\label{eq:exp_asr_ue_closed}
\mathbb{E}[\mathrm{ASR}]_{\mathrm{UE}}
= \frac{1}{k}\Big[(1-p)(1-q)^{k-1} + \frac{p}{q}\big(1-(1-q)^{k}\big)\Big].
\end{equation}
For \(q\to 0\), the second term tends to \(\tfrac{p}{k}\cdot k = p\), as expected.

\subsubsection{SHE Protocol} \label{app:asr_she}

The expected ASR of SHE is defined as the probability that \( \hat{x} = x \):
\begin{equation*}
    \mathbb{E}[\text{ASR}]_{\mathrm{SHE}} = \Pr[\hat{x} = x] = \Pr\left [ y_x > \max_{i \neq x} y_i \right] \mathrm{.}
\end{equation*}

Define the random variables:
\begin{align*}
    y_x &= 1 + Z_x, \quad \text{where } Z_x \sim \text{Laplace}(0, b)\mathrm{,} \\
    y_i &= 0 + Z_i = Z_i, \quad \text{where } Z_i \sim \text{Laplace}(0, b)\mathrm{,} \quad \forall i \neq x
\end{align*}

All \( Z_i \) and \( Z_x \) are independent random variables.
Let \( M = \max_{i \neq x} y_i = \max_{i \neq x} Z_i \). 
Then, the expected ASR becomes:
\begin{equation*}
    \begin{aligned}
    \mathbb{E}[\text{ASR}]_{\mathrm{SHE}} &= \Pr\left[ y_x > M \right]\\ 
    &= \Pr\left[ 1 + Z_x > M \right] \\
    &= \Pr\left[ Z_x > M - 1 \right] \mathrm{.}
    \end{aligned}
\end{equation*}

The cumulative distribution function (CDF) of the Laplace distribution \( Z_x \sim \text{Laplace}(0, b) \) is:
\begin{equation*}
    F_Z(z) = 
    \begin{cases}
        \dfrac{1}{2} \exp\left[ \dfrac{z}{b} \right], & \text{if } z \leq 0  \mathrm{,}\\
        1 - \dfrac{1}{2} \exp\left[ -\dfrac{z}{b} \right], & \text{if } z > 0  \mathrm{.}
    \end{cases}
\end{equation*}

The probability density function (PDF) of \( Z_x \) is:
\begin{equation*}
    f_Z(z) = \dfrac{1}{2b} \exp\left[ -\dfrac{|z|}{b} \right] \mathrm{.}
\end{equation*}

For \( M = \max_{i \neq x} Z_i \), since \( Z_i \) are independent and identically distributed (i.i.d.), the CDF of \( M \) is:
\begin{equation}
    F_M(m) = \left[ F_Z(m) \right]^{k - 1} \mathrm{.}
\end{equation}

The PDF of \( M \) is then:
\begin{equation}
    f_M(m) = (k - 1) \left[ F_Z(m) \right]^{k - 2} f_Z(m) \mathrm{.}
\end{equation}

The ASR can be expressed as:
\begin{equation}
\mathbb{E}[\text{ASR}]_{\mathrm{SHE}} = \int_{-\infty}^{\infty} \Pr\left[ Z_x > m - 1 \right] f_M(m) \, dm  \mathrm{.}
\end{equation}

Since \( Z_x \) and \( M \) are independent, \( \Pr\left[ Z_x > m - 1 \right] = 1 - F_Z(m - 1) \). Therefore:
\begin{equation} \label{eq:full_exp_asr_she}
    \mathbb{E}[\text{ASR}]_{\mathrm{SHE}} = \int_{-\infty}^{\infty} \left[ 1 - F_Z(m - 1) \right] f_M(m) \, dm  \mathrm{.}
\end{equation}

\textbf{Empirical Estimation via Simulation.} In this work, we estimate the expected ASR in Equation~\eqref{eq:full_exp_asr_she} empirically using Monte Carlo simulations, following:

\begin{enumerate}
    \item \textbf{Generate Samples}:
    \begin{itemize}
        \item Sample \( Z_x \) from \( \text{Laplace}(0, b) \).
        \item Sample \( Z_i \) for \( i \neq x \) and compute \( M = \max_{i \neq x} Z_i \).
    \end{itemize}
    \item \textbf{Compute Success Indicator}:
    \begin{itemize}
        \item For each sample, check if \( 1 + Z_x > M \).
    \end{itemize}
    \item \textbf{Estimate ASR}:
    \begin{itemize}
        \item The ASR is estimated as the proportion of times \( 1 + Z_x > M \) holds over all samples.
    \end{itemize}
\end{enumerate}

\subsubsection{THE Protocol} \label{app:asr_the}
Following the attack strategy for THE in Section~\ref{sub:the_protocol}, we have:

\begin{itemize}
    \item \textbf{Event 0:} The bit corresponding to the user's value \( x \) is less than \(\theta\) (\ie, remains 0) and all other bits also remain 0.
    \begin{itemize}
        \item The attacker's guess is uniformly distributed over all \( k \) possible values.
       
       \item Success rate: \(\frac{1}{k}\).
       
       \item Probability: \(\Pr(\text{Event 0}) = (1 - p)(1 - q)^{k-1}\).
    \end{itemize}

    \item \textbf{Event 1:} The bit corresponding to the user's value \( x \) is greater than \(\theta\) (\ie, flips to 1) and \(m-1\) other bits also flip to 1.
    \begin{itemize}
        \item The attacker's guess is uniformly distributed over the bits set to 1.
        
        \item If \(m\) bits are set to 1, the success rate is \(\frac{1}{m}\)
        
        \item Probability: \(\Pr(\text{Event 1 with } m \text{ bits set to 1}) = \binom{k-1}{m-1} p (q)^{m-1} (1 - q)^{k-m}\).
    \end{itemize}
\end{itemize}

Thus, by combining these two events, we can derive the expected ASR as:
\[
\begin{aligned}
    \mathbb{E}[\text{ASR}]_{\mathrm{THE}} =& \Pr(\text{Event 0}) \cdot \frac{1}{k} \\
    &+ \sum_{m=1}^{k} \Pr(\text{Event 1 with } m \text{ bits set to 1}) \cdot \frac{1}{m} \mathrm{.}
\end{aligned}
\]

More formally, the probability calculations of each event are:

\begin{enumerate}
    \item \textbf{Probability of Event 0:} 
    \[
        \Pr(\text{Event 0}) = (1 - p)(1 - q)^{k-1} \mathrm{.}
    \]

    \item \textbf{Probability of Event 1 with \( m \) bits set to 1:} 
    \[
    \begin{aligned}
        \Pr(\text{Event 1 with } m \text{ bits set to 1}) =& \\
        \binom{k-1}{m-1} p (q)^{m-1} (1 - q)^{k-m} \mathrm{.} &
    \end{aligned}
    \]
\end{enumerate}

Combining these probabilities, the expected ASR for THE is:
\[
\begin{aligned}
    \mathbb{E}[\text{ASR}]_{\mathrm{THE}} &= (1 - p)(1 - q)^{k-1} \cdot \frac{1}{k} \\
    &+ \sum_{m=1}^{k} \binom{k-1}{m-1} p (q)^{m-1} (1 - q)^{k-m} \cdot \frac{1}{m} \mathrm{.}
\end{aligned}
\]

Using $X \sim \mathrm{Binomial}(k-1, q)$ where $M = X + 1$, we apply:
\[
\mathbb{E}\left[\frac{1}{X+1}\right] = \frac{1 - (1-q)^k}{kq} \mathrm{.}
\]

Final closed-form expression:
\begin{equation} \label{eq:exp_asr_the_closed}
    \mathbb{E}[\mathrm{ASR}]_{\mathrm{THE}} = (1 - p)\frac{(1-q)^{k-1}}{k} + \frac{p}{kq}\left(1 - (1-q)^k\right) \mathrm{.}
\end{equation}

\subsection{Optimization Strategies for Parameter Selection}
\label{app:optimization_methods}

This section provides a detailed overview of the multi-objective optimization strategies supported by our framework for selecting protocol parameters in adaptive LDP mechanisms.
Given a Pareto frontier generated from exhaustive grid search~\cite{bergstra2012random}, we apply six classical selection methods to identify the most suitable trade-off between Attacker Success Rate (ASR) and Mean Squared Error (MSE), as described in Section~\ref{sub:asr_mse_framework}. 
Each strategy interprets the frontier differently and is designed to meet distinct design goals~\cite{Donoso2007}. 

\begin{itemize}

    \item \textbf{Utopia Point Minimization}:  
    Selects the Pareto point closest (in Euclidean distance) to the theoretical ideal point \((0, 0)\), representing zero ASR and zero MSE.  
    This method emphasizes proximity to perfect privacy and utility, making it a conceptually intuitive baseline.

    \item \textbf{Weighted Scalarization (Default)}:  
    Computes a scalarized objective as in Equation~\eqref{eq:objective_function}, using user-defined weights \((w_{\text{ASR}}, w_{\text{MSE}})\).
    The point that minimizes this weighted sum is selected.  
    This strategy is tunable, interpretable, and easy to implement, making it the default in our framework.
    As mentioned in the main paper, we set \(w_{\text{ASR}}=w_{\text{MSE}}=0.5\) by default.

    \item \textbf{Elbow (Knee) Method}:  
    Identifies the point on the frontier that exhibits the greatest curvature, corresponding to the location where further reductions in MSE yield diminishing returns in terms of ASR.  
    It does so by maximizing the perpendicular distance from the frontier to the chord connecting the extreme points.

    \item \textbf{$\boldsymbol{\epsilon}$-Constraint Selection}:  
    Filters the Pareto frontier to retain only points with ASR below a predefined threshold \(\epsilon_{\text{ASR}}\), and among them selects the one with the lowest MSE.  
    If no feasible point exists, the strategy defaults to the utopia point.  
    This is particularly useful in privacy-critical settings where strict ASR bounds are imposed.
    We set \(\epsilon_{\text{ASR}}=0.1\) by default.

    \item \textbf{Hypervolume Contribution}:  
    Selects the Pareto point that contributes the most area (in 2D) to the dominated space relative to a reference point.  
    This method emphasizes diversity and coverage of the solution set, making it attractive when the overall Pareto frontier spread is important.
    We set the reference point as \((1.0, 1.0)\).

    \item \textbf{Augmented Tchebycheff Scalarization}:  
    Minimizes the augmented Chebyshev objective: $\max_i \{ w_i f_i \} + \rho \sum_i w_i f_i$, where \(f = (\text{ASR}, \text{MSE})\), \(w_i\) are normalized weights, and \(\rho\) is a small regularization factor.
    This approach balances worst-case performance and average cost, useful when robustness is a key concern.
    We set $w_{\text{ASR}}=w_{\text{MSE}}=0.5$ and $\rho=10^{-6}$ by default.

\end{itemize}

\begin{figure*}[!htb]
    \centering
    \begin{subfigure}[t]{\textwidth}
        \centering
        \includegraphics[width=\linewidth]{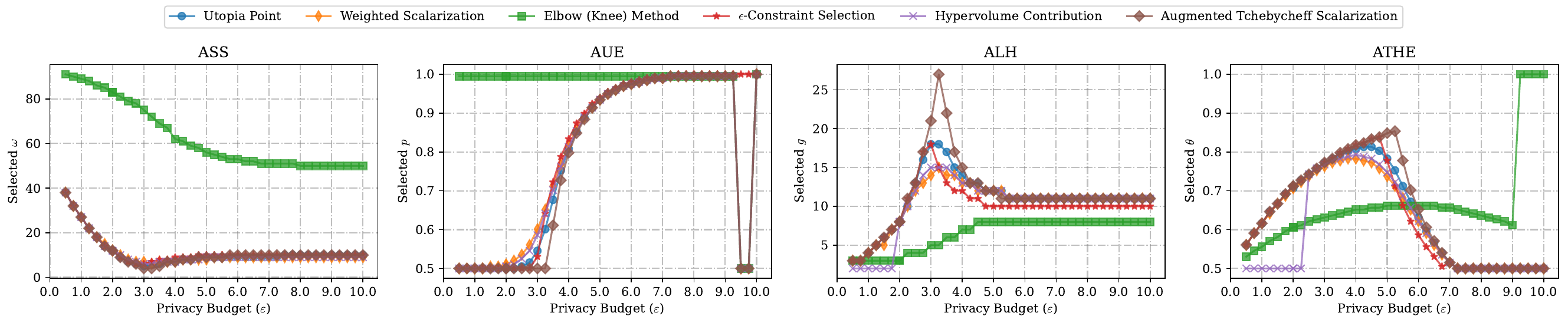}
        \caption{Domain size \(k = 100\).}
        \label{fig:optimization_methods_k_100}
    \end{subfigure}
    
    \vspace{0.5cm} 

    \begin{subfigure}[t]{\textwidth}
        \centering
        \includegraphics[width=\linewidth]{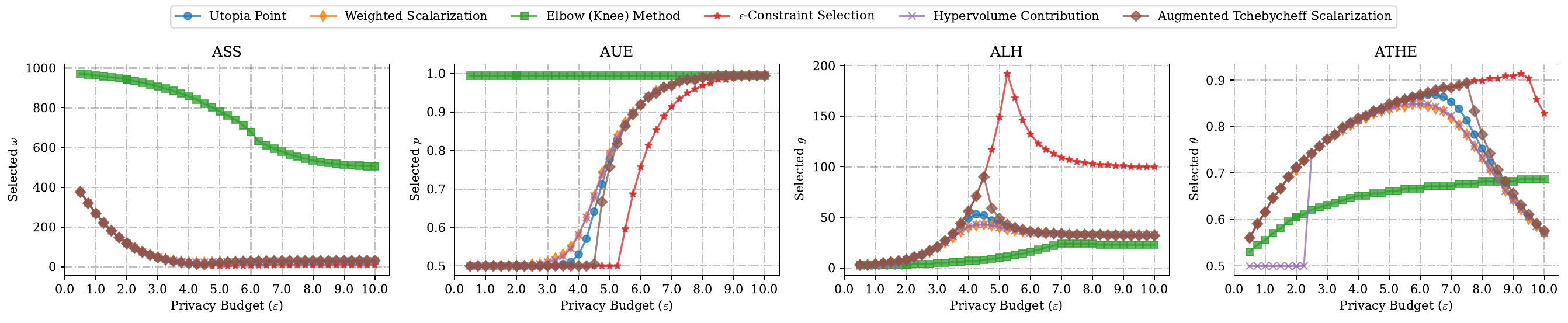}
        \caption{Domain size \(k = 1000\).}
        \label{fig:optimization_methods_k_1000}
    \end{subfigure}
    
    \vspace{0.5cm} 

    \begin{subfigure}[t]{\textwidth}
        \centering
        \includegraphics[width=\linewidth]{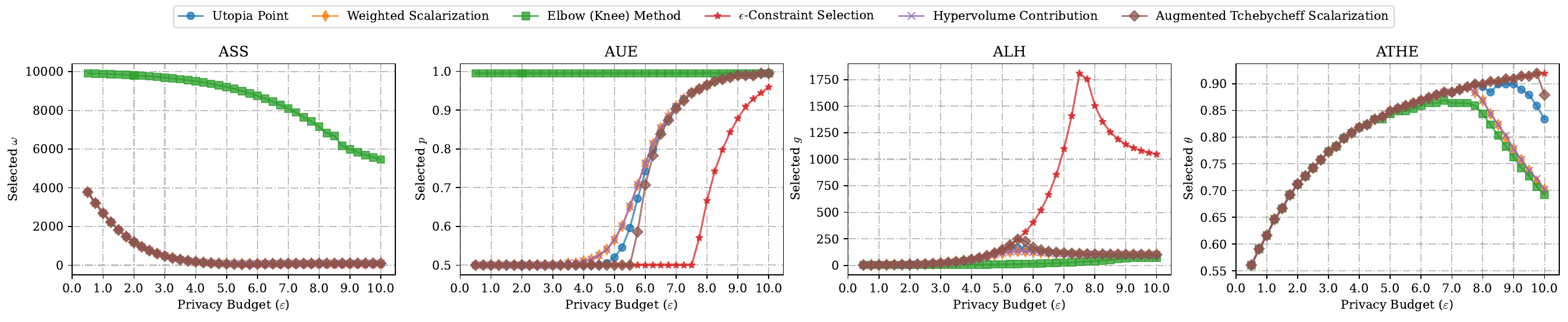}
        \caption{Domain size \(k = 10000\).}
        \label{fig:optimization_methods_k_10000}
    \end{subfigure}
    
    \caption{Selected protocol parameters across privacy budgets $\varepsilon$ for each adaptive mechanism (ASS, AUE, ALH, ATHE), under six multi-objective selection strategies. 
    Each subplot considers a range of privacy budgets $\varepsilon \in (0.5, 10)$ and varying domain sizes \(k \in \{100, 1000, 10000\}\).
    Each curve shows how a strategy (\eg, Utopia Point, Weighted Scalarization, Elbow Method) selects optimal parameters from the Pareto frontier of (ASR, MSE) trade-offs.
    While most strategies converge to similar configurations, the Elbow method exhibits distinctive behavior due to its curvature-based criterion.}
    \label{fig:appendix_optimization}
\end{figure*}

Figure~\ref{fig:appendix_optimization} presents the parameters selected by each strategy across a range of $\varepsilon$ values and domain sizes \(k \in \{100, 1000, 10000\}\) (representing small, medium, and large domains).
These plots provide insights into the behavior and consistency of the selection methods.
We observe that most strategies tend to select similar parameter values, particularly in the high-privacy and low-privacy regimes.
In contrast, the Elbow method often diverges from the others due to its reliance on geometric curvature, which can result in unstable behavior across different settings.
Overall, \textbf{Weighted Scalarization} offers a robust compromise, simple to implement, interpretable and tunable, making it the \textbf{default strategy} in our implementation.
Nevertheless, our implementation supports all strategies, enabling practitioners to select methods that best align with their specific privacy, utility or robustness requirements.

\subsection{Impact of Weights in the Objective Function} \label{app:weights_results}

\begin{figure*}[htb]
    \centering
    \includegraphics[width=0.72\linewidth]{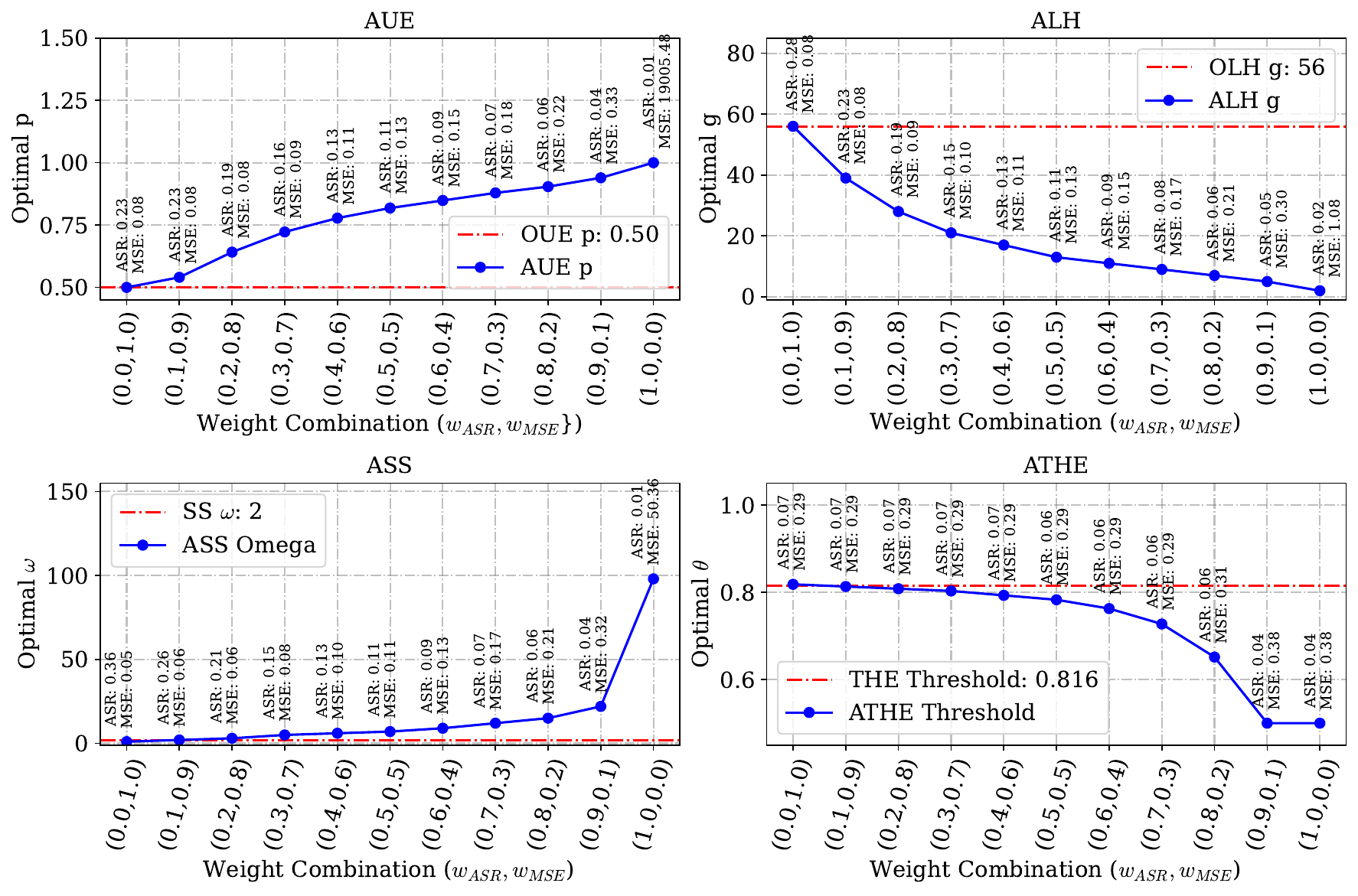}
    \caption{%
    Optimal parameter choices for each adaptive protocol as a function of the weight combination \((w_{\text{ASR}}, w_{\text{MSE}})\), evaluated at \(k=100\) and \(\varepsilon=4\). 
    Each sub-figure compares the adaptive protocol's chosen parameter (\textcolor{blue}{blue color} curve) against the corresponding parameter choice in the original, non-adaptive protocol (\textcolor{red}{red color} dashed line). 
    }
    \label{fig:weights_impact}
\end{figure*}

Thus far, we have focused on analyzing the performance of adaptive protocols under fixed objective configurations for the weights $(w_{\text{ASR}} = w_{\text{MSE}} = 0.5)$ in Eq.~\eqref{eq:objective_function}. 
However, our proposed multi-objective framework introduces a new degree of freedom: practitioners can adjust the relative importance of ASR \vs{} variance (MSE) when optimizing LDP protocols. 
Figure~\ref{fig:weights_impact} illustrates how varying these weight combinations \((w_{\text{ASR}}, w_{\text{MSE}})\) influences the ASR, MSE, and optimal parameter choices for AUE (\(p\)), ALH (\(g\)), ASS (\(\omega\)) and ATHE (\(\theta\)) under a fixed setting (\(k=100\), \(\varepsilon=4\)).

From Figure~\ref{fig:weights_impact}, one can notice that as the weight on ASR $w_{\text{ASR}}$ increases, each adaptive protocol tends to choose parameter values that more aggressively reduce the ASR at the cost of increasing the MSE. 
Conversely, placing greater emphasis on MSE drives parameters toward configurations closer to or equal to those of the original protocols (\ie, SS, OUE, OLH and THE), aiming to preserve utility even if it elevates the ASR.
These results confirm the benefits of our two-objective optimization framework: rather than a static parameter choice, practitioners can tune the protocol parameters in response to changing priorities, achieving a more flexible trade-off between privacy (ASR) and utility (MSE).

\subsection{Adaptive and Optimized Parameters} \label{app:parameter_results}

\begin{figure*}[!htb]
    \centering
    \includegraphics[width=0.72\linewidth]{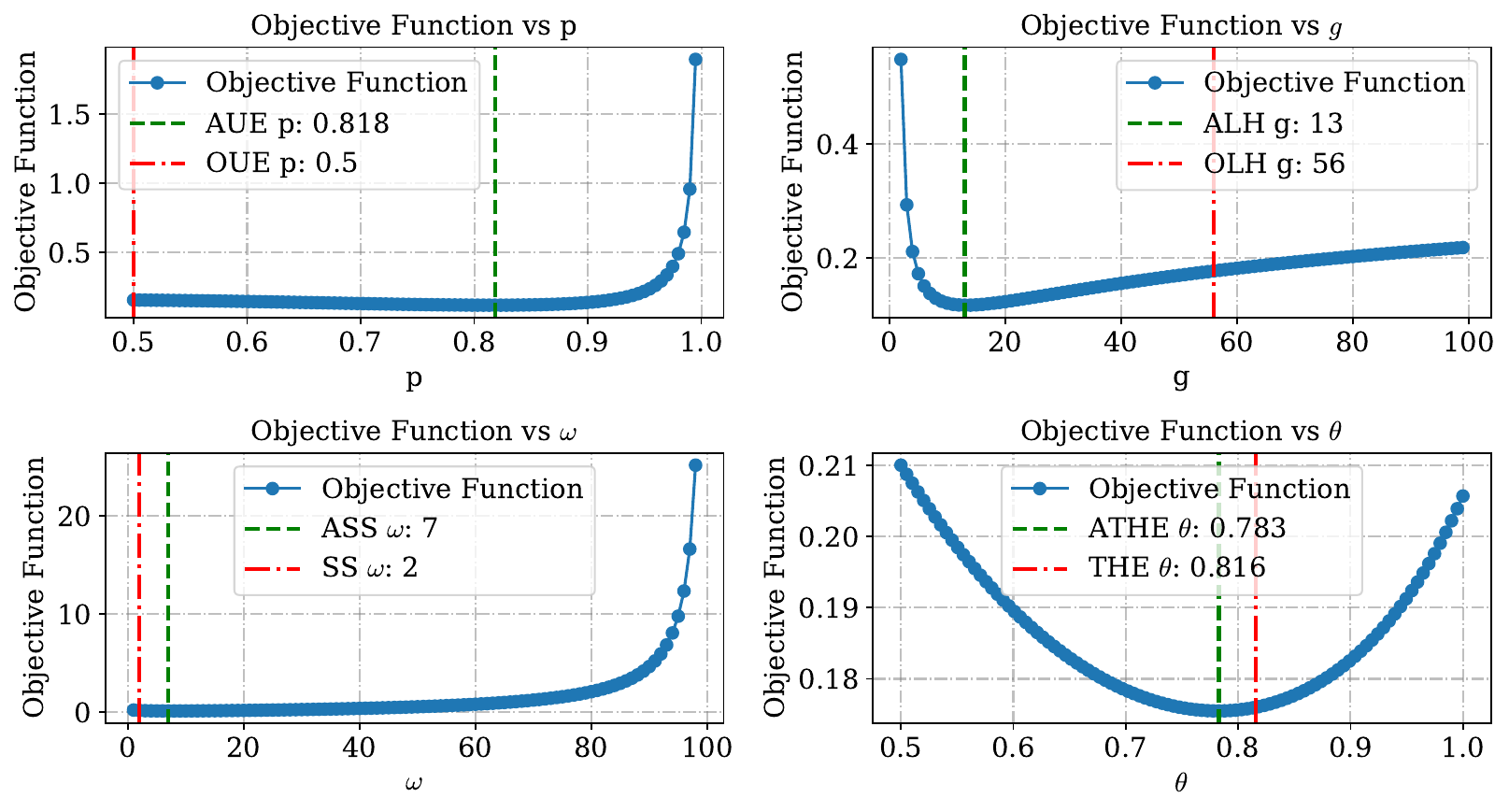}
    \caption{Objective function value as a function of key parameters for our adaptive protocols (AUE, ALH, ASS and ATHE) compared with their state-of-the-art counterparts (OUE, OLH, SS and THE). Vertical dashed lines (\textcolor{mygreen}{green color}) indicate the optimal parameter values selected by our adaptive protocols, while vertical dash-dotted lines (\textcolor{red}{red color}) represent the fixed parameter values of the state-of-the-art protocols.
    Without loss of generality, we set $k=100$ and $\varepsilon=4$.}
    \label{fig:parameter_analysis}
\end{figure*}

In this section, we analyze the optimization of parameters in our adaptive protocols (AUE, ALH, ASS and ATHE) by examining the behavior of their objective functions (Equations~\eqref{eq:optimization_ass}--\eqref{eq:optimization_athe}), which balance the ASR-MSE trade-off. 
Figure~\ref{fig:parameter_analysis} illustrates the objective function as a function of key parameters: AUE (\(p\)), ALH (\(g\)), ASS (\(\omega\)) and ATHE (\(\theta\)), with a fixed $k=100$ and $\varepsilon=4$. 
The selected parameters for our adaptive protocols are compared against the fixed state-of-the-art parameters of OUE, OLH, SS and THE.

For \textbf{AUE}, we observe in the top-left plot that the objective function reaches its minimum at \(p = 0.818\), which is notably higher than the fixed \(p = 0.5\) used by OUE. 
This also means that parameter $q$ will increase to satisfy $\varepsilon$-LDP, \ie, increasing the probability of reporting random bits.
For \textbf{ALH}, as shown in the top-right plot, the optimal hash domain size is reduced to \(g=13\)  in our adaptive protocol compared to \(g = 56\) in OLH. 
This reduction in \(g\) lowers the ASR at the expense of slightly increased variance, aligning with our adaptive objective to achieve a better balance between privacy and utility.
For \textbf{ASS}, as depicted in the bottom-left plot, the optimal subset size \(\omega = 7\) contrasts with the fixed \(\omega = 2\) used in SS. 
The larger \(\omega\) effectively spreads the probability mass across a larger subset, reducing ASR while incurring a higher variance. 
For \textbf{ATHE}, the bottom-right plot shows that the adaptive threshold \(\theta = 0.783\) is slightly lower than the fixed threshold \(\theta = 0.816\) in THE. 
This subtle adjustment enables ATHE to reduce privacy attacks while maintaining competitive variance levels, showcasing the precision of our adaptive optimization.

\begin{leftbar}
\noindent \textbf{Adaptive protocols optimize parameters for better trade-offs:} 
Our findings demonstrate that the optimized parameters selected by adaptive protocols significantly differ from the fixed parameters of state-of-the-art protocols, resulting in improved robustness to privacy attacks with controlled increases in variance. 
This optimization highlights the effectiveness of our adaptive mechanisms in better navigating the privacy-utility trade-off space.
\end{leftbar}

\subsection{Additional Analytical Results for the ASR \vs{} MSE Trade-Off} \label{app:add_asr_mse_tradeoff}

\begin{figure*}[!ht]
    \centering
    \includegraphics[width=1\linewidth]{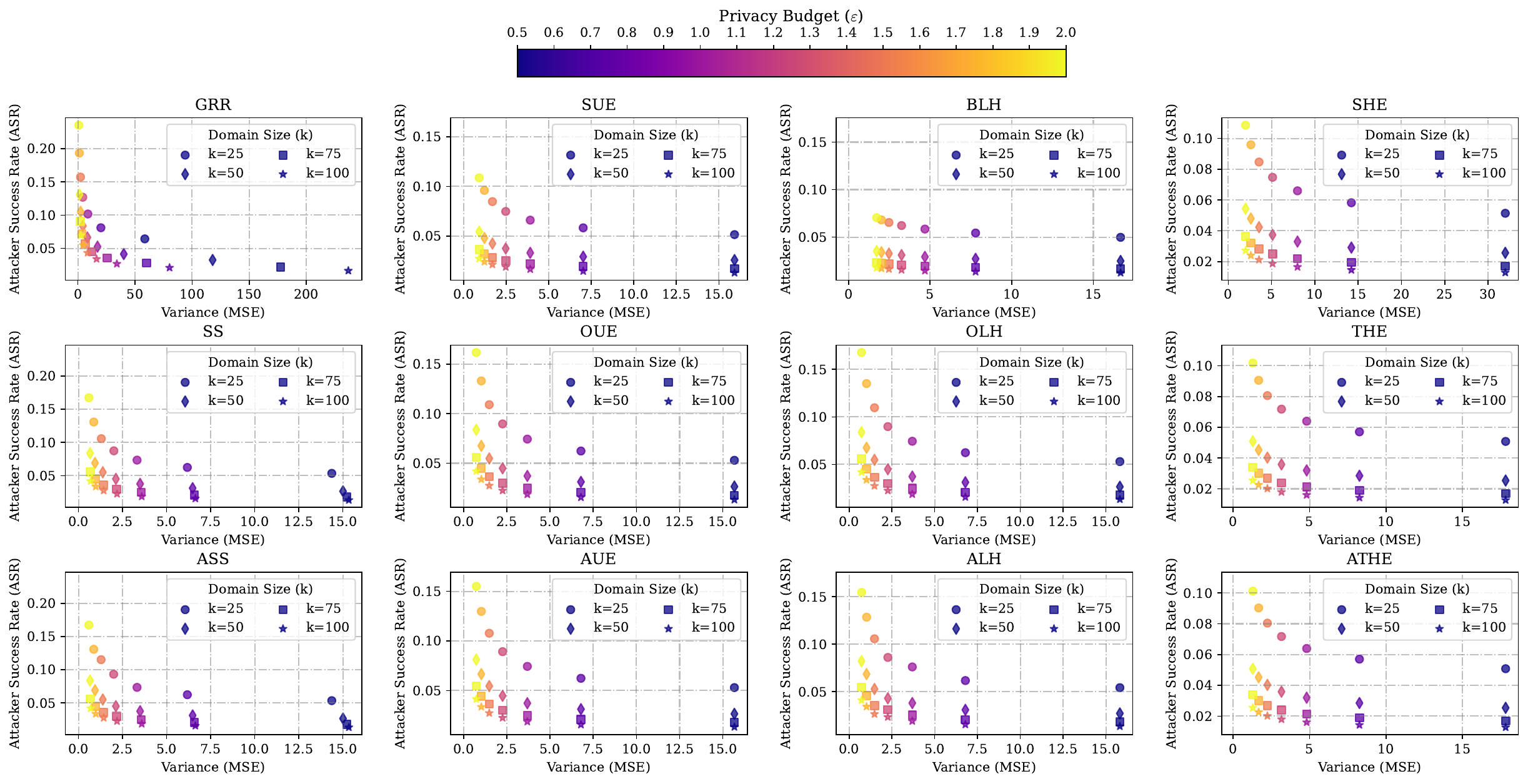}
    \caption{Attacker Success Rate (ASR) \vs{} Variance (MSE) for numerous LDP frequency estimation protocols. 
    Each plot shows how each protocol performs under varying privacy budgets $\varepsilon$ and domain sizes ($k$), illustrating the trade-off between adversarial success rate (ASR) and utility (MSE). 
    State-of-the-art LDP protocols (\ie, GRR, SUE, BLH, SHE, SS, OUE, OLH, and THE) are compared against our adaptive counterparts (\ie, ASS, AUE, ALH, and ATHE). 
    Each point represents a different configuration of $\varepsilon$ (in high privacy regimes) and $k$ (small domain), with colors indicating the privacy budget level.}
    \label{fig:asr_vs_mse_analysis_high_priv_small_domain}
\end{figure*}

\begin{figure*}[t]
    \centering
    \includegraphics[width=1\linewidth]{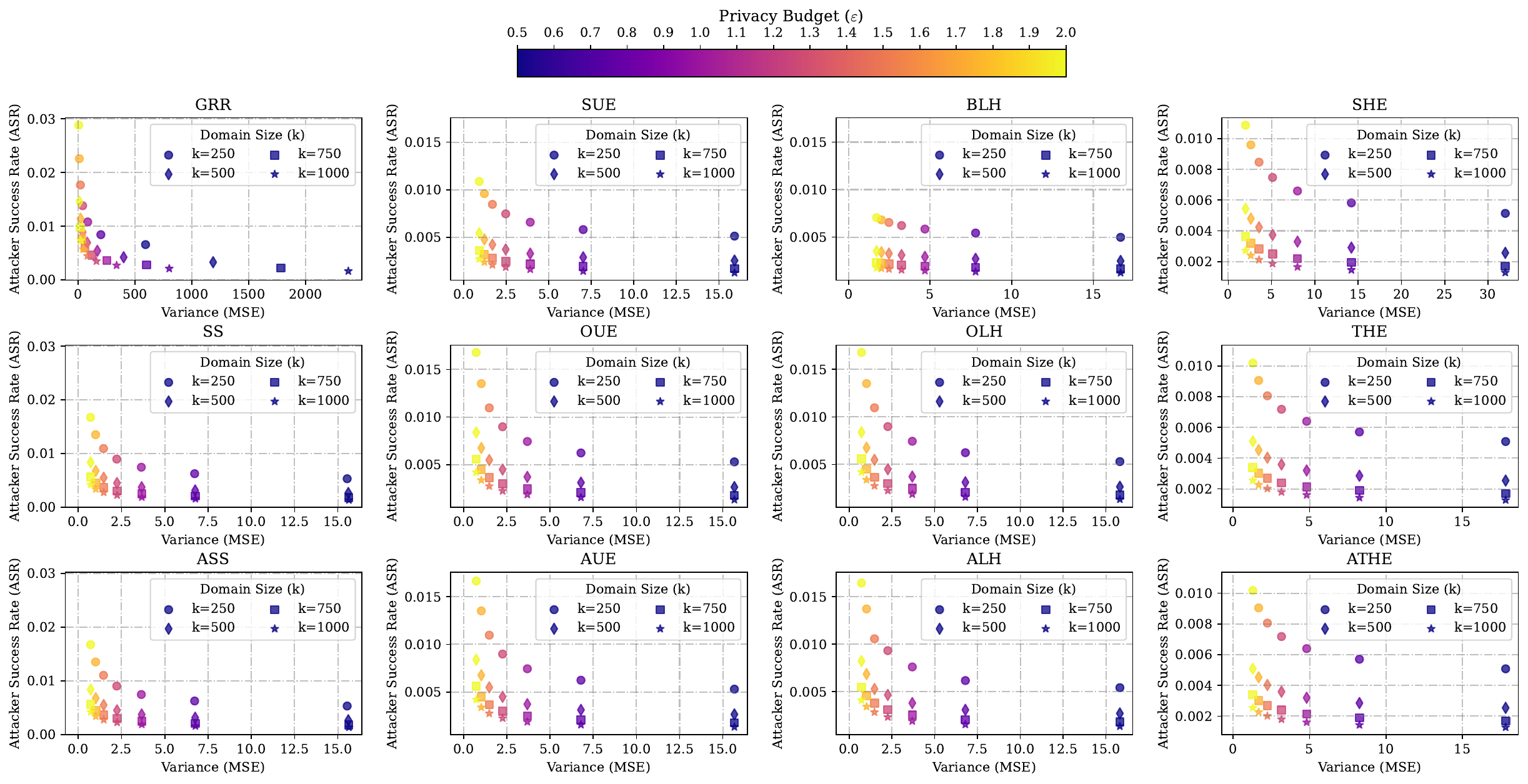}
    \caption{Attacker Success Rate (ASR) \vs{} Variance (MSE) for numerous LDP frequency estimation protocols. 
    Each plot shows how each protocol performs under varying privacy budgets $\varepsilon$ and domain sizes ($k$), illustrating the trade-off between adversarial success rate (ASR) and utility (MSE). 
    State-of-the-art LDP protocols (\ie, GRR, SUE, BLH, SHE, SS, OUE, OLH, and THE) are compared against our adaptive counterparts (\ie, ASS, AUE, ALH, and ATHE). 
    Each point represents a different configuration of $\varepsilon$ (in high privacy regimes) and $k$ (medium domain), with colors indicating the privacy budget level.}
    \label{fig:asr_vs_mse_analysis_high_priv_medium_domain}
\end{figure*}

\begin{figure*}[t]
    \centering
    \includegraphics[width=1\linewidth]{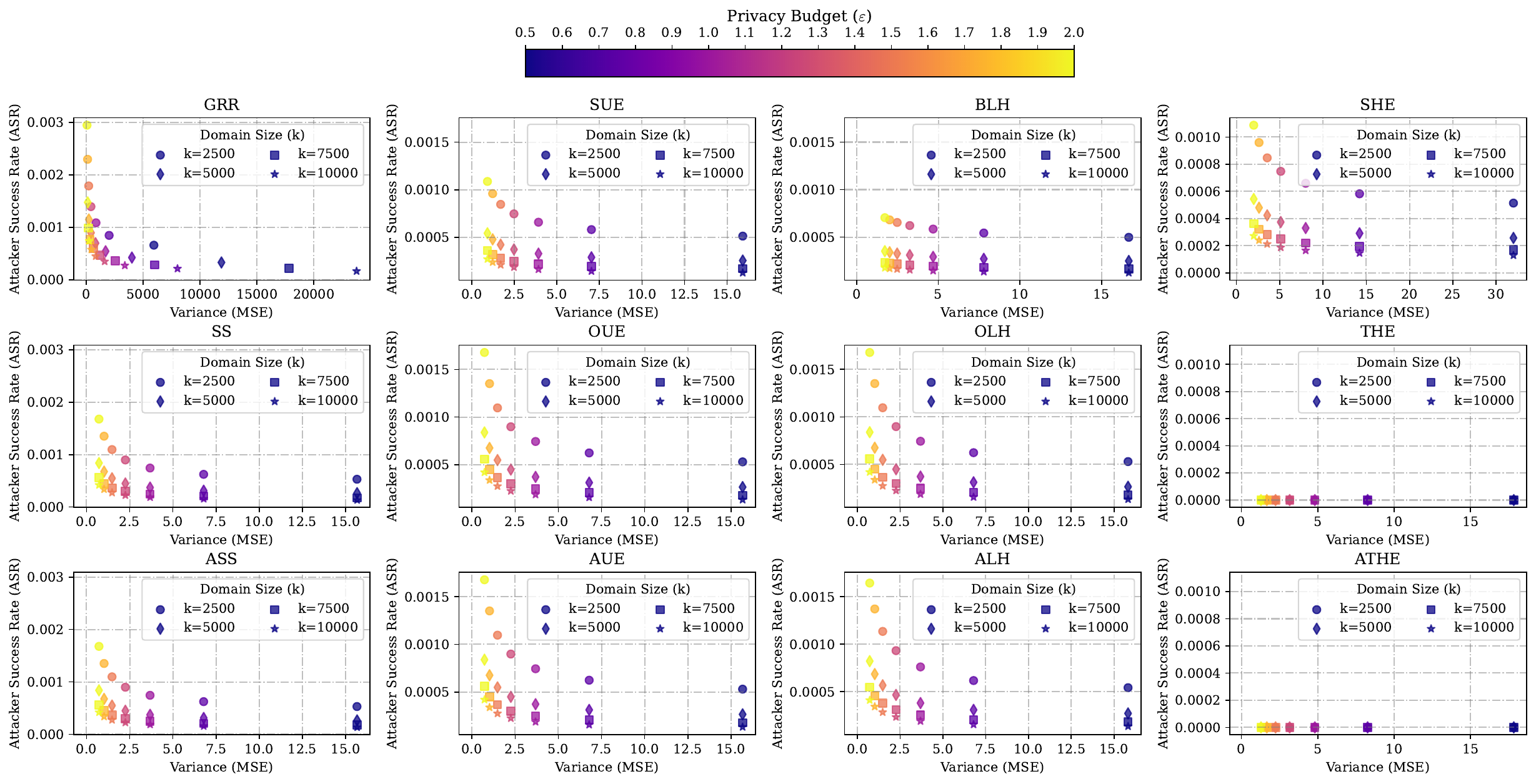}
    \caption{Attacker Success Rate (ASR) \vs{} Variance (MSE) for numerous LDP frequency estimation protocols. 
    Each plot shows how each protocol performs under varying privacy budgets $\varepsilon$ and domain sizes ($k$), illustrating the trade-off between adversarial success rate (ASR) and utility (MSE). 
    State-of-the-art LDP protocols (\ie, GRR, SUE, BLH, SHE, SS, OUE, OLH, and THE) are compared against our adaptive counterparts (\ie, ASS, AUE, ALH, and ATHE). 
    Each point represents a different configuration of $\varepsilon$ (in high privacy regimes) and $k$ (large domain), with colors indicating the privacy budget level.}
    \label{fig:asr_vs_mse_analysis_high_priv_large_domain}
\end{figure*}

\begin{figure*}[t]
    \centering
    \includegraphics[width=1\linewidth]{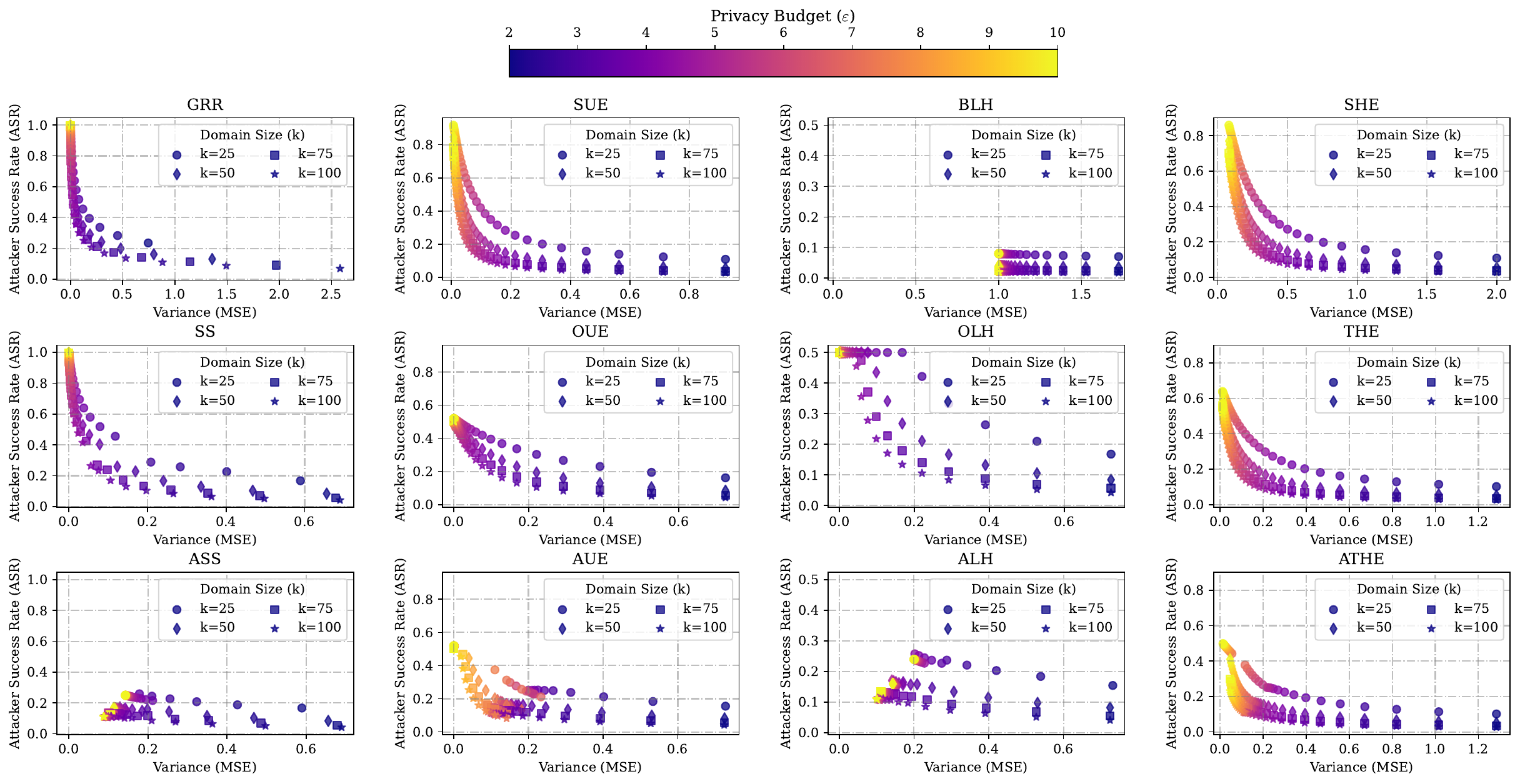}
    \caption{Attacker Success Rate (ASR) \vs{} Variance (MSE) for numerous LDP frequency estimation protocols. 
    Each plot shows how each protocol performs under varying privacy budgets $\varepsilon$ and domain sizes ($k$), illustrating the trade-off between adversarial success rate (ASR) and utility (MSE). 
    State-of-the-art LDP protocols (\ie, GRR, SUE, BLH, SHE, SS, OUE, OLH, and THE) are compared against our adaptive counterparts (\ie, ASS, AUE, ALH, and ATHE). 
    Each point represents a different configuration of $\varepsilon$ (in medium to low privacy regimes) and $k$ (medium domain), with colors indicating the privacy budget level.}
    \label{fig:asr_vs_mse_analysis_low_priv_small_domain}
\end{figure*}

\begin{figure*}[t]
    \centering
    \includegraphics[width=1\linewidth]{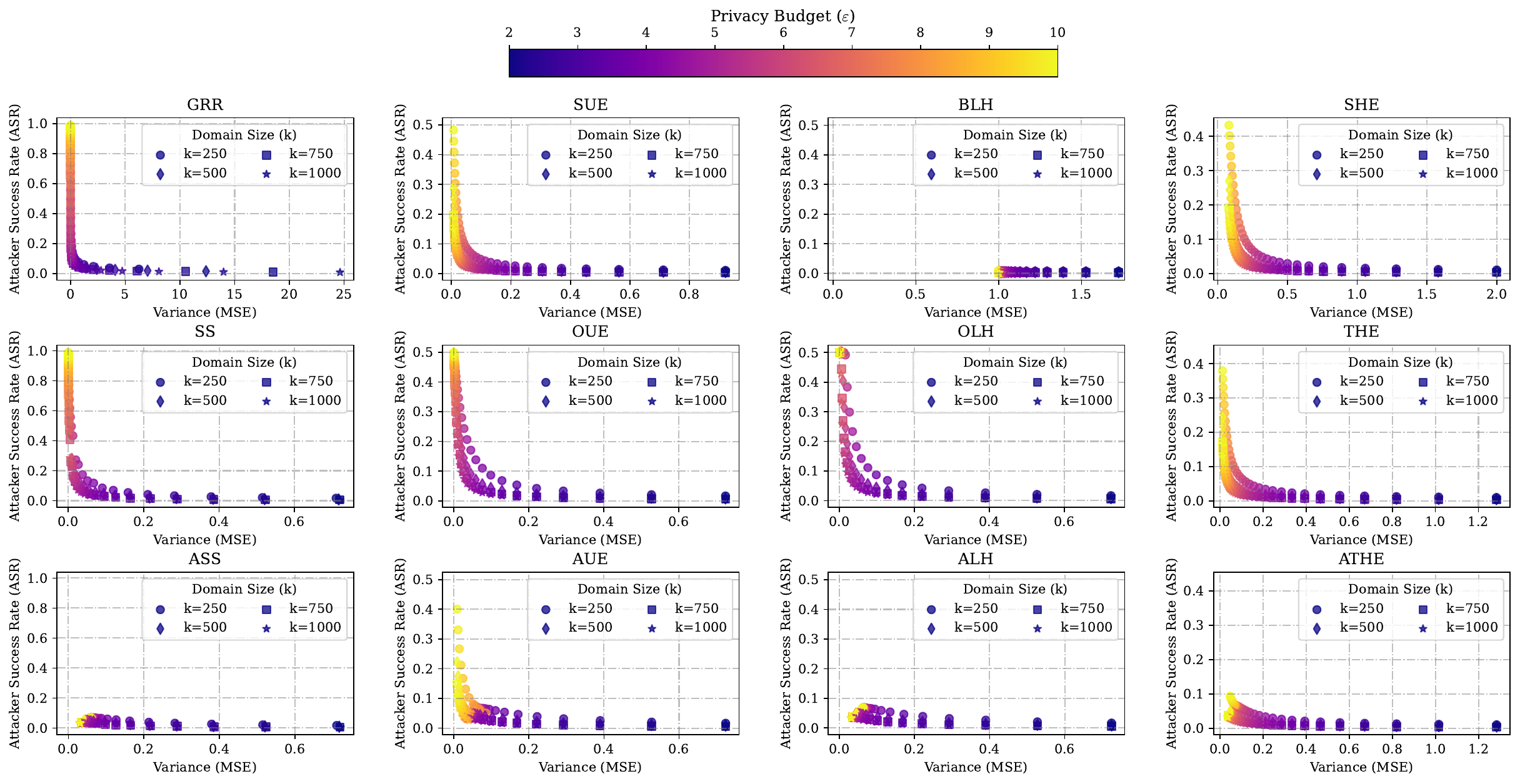}
    \caption{Attacker Success Rate (ASR) \vs{} Variance (MSE) for numerous LDP frequency estimation protocols. 
    Each plot shows how each protocol performs under varying privacy budgets $\varepsilon$ and domain sizes ($k$), illustrating the trade-off between adversarial success rate (ASR) and utility (MSE). 
    State-of-the-art LDP protocols (\ie, GRR, SUE, BLH, SHE, SS, OUE, OLH, and THE) are compared against our adaptive counterparts (\ie, ASS, AUE, ALH, and ATHE). 
    Each point represents a different configuration of $\varepsilon$ (in medium to low privacy regimes) and $k$ (medium domain), with colors indicating the privacy budget level.}
    \label{fig:asr_vs_mse_analysis_low_priv_medium_domain}
\end{figure*}

\begin{figure*}[t]
    \centering
    \includegraphics[width=1\linewidth]{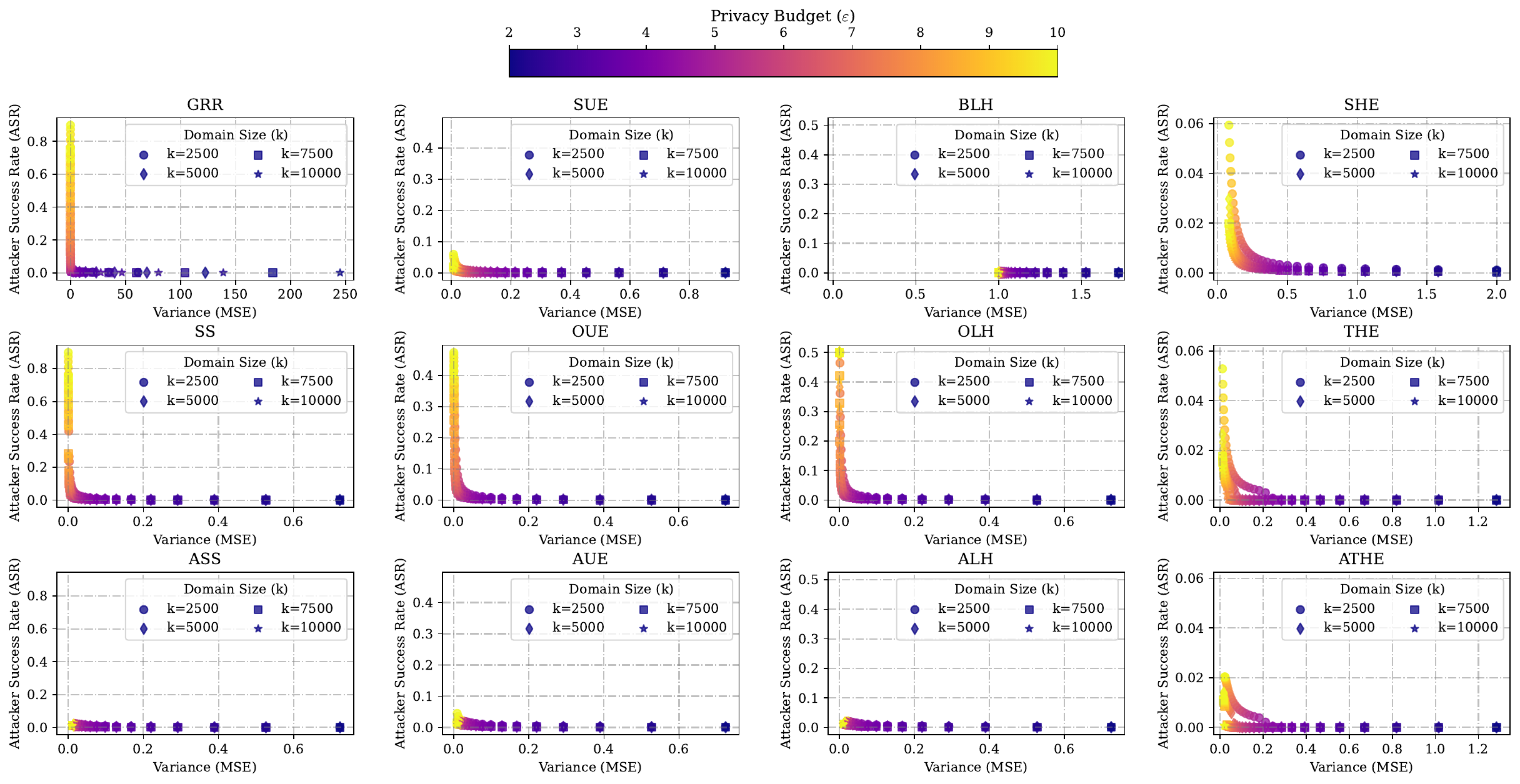}
    \caption{Attacker Success Rate (ASR) \vs{} Variance (MSE) for numerous LDP frequency estimation protocols. 
    Each plot shows how each protocol performs under varying privacy budgets $\varepsilon$ and domain sizes ($k$), illustrating the trade-off between adversarial success rate (ASR) and utility (MSE). 
    State-of-the-art LDP protocols (\ie, GRR, SUE, BLH, SHE, SS, OUE, OLH, and THE) are compared against our adaptive counterparts (\ie, ASS, AUE, ALH, and ATHE). 
    Each point represents a different configuration of $\varepsilon$ (in medium to low privacy regimes) and $k$ (large domain), with colors indicating the privacy budget level.}
    \label{fig:asr_vs_mse_analysis_low_priv_large_domain}
\end{figure*}

To complement the results of Section~\ref{sub:asr_mse_results}, Figures~\ref{fig:asr_vs_mse_analysis_high_priv_small_domain} to~\ref{fig:asr_vs_mse_analysis_low_priv_large_domain} illustrate the ASR-MSE trade-off considering:

\begin{itemize}
    \item Figure~\ref{fig:asr_vs_mse_analysis_high_priv_small_domain}: high privacy regime and small domain size.

    \item Figure~\ref{fig:asr_vs_mse_analysis_high_priv_medium_domain}: high privacy regime and medium domain size.

    \item Figure~\ref{fig:asr_vs_mse_analysis_high_priv_large_domain}: high privacy regime and large domain size.

    \item Figure~\ref{fig:asr_vs_mse_analysis_low_priv_small_domain}: medium to low privacy regimes and small domain size.

    \item Figure~\ref{fig:asr_vs_mse_analysis_low_priv_medium_domain}: medium to low privacy regimes and medium domain size.

    \item Figure~\ref{fig:asr_vs_mse_analysis_low_priv_large_domain}: medium to low privacy regimes and large domain size.
\end{itemize}

\end{document}